\documentclass{aastex631}

\usepackage{amsmath}

\usepackage{hyperref}
\usepackage[nameinlink,capitalise,noabbrev]{cleveref}

\usepackage{tikz}
\usetikzlibrary{shapes.geometric, arrows}
\tikzstyle{startstop} = [rectangle, rounded corners, 
minimum width=1.5cm, 
minimum height=0.7cm,
text centered, 
draw=black, 
fill=white]

\tikzstyle{io} = [trapezium, 
trapezium stretches=true, 
trapezium left angle=70, 
trapezium right angle=110, 
minimum width=2cm, 
minimum height=0.7cm, text centered, 
draw=black, fill=white]

\tikzstyle{process} = [rectangle, 
minimum width=1.5cm, 
minimum height=0.7cm, 
text centered, 
text width=3cm, 
draw=black, 
fill=white]

\tikzstyle{decision} = [diamond, 
aspect=3,
minimum width=2cm, 
minimum height=0.7cm, 
text centered, 
draw=black, 
fill=white]
\tikzstyle{arrow} = [thick,->,>=stealth]

\usepackage{soul}
\usepackage[nolist]{acronym} 

\begin{document}

\title{A Fast, Second-Order Accurate Poisson Solver in Spherical Polar Coordinates}

\author{Jeonghyeon Ahn}
\affiliation{Department of Physics \& Astronomy, Seoul National University, Seoul 08826, Korea}

\author[0000-0003-4625-229X]{Woong-Tae Kim}
\affiliation{Department of Physics \& Astronomy, Seoul National University, Seoul 08826, Korea}
\affiliation{SNU Astronomy Research Center, Seoul National University, 1 Gwanak-ro, Gwanak-gu, Seoul 08826, Republic of Korea}

\author[0000-0003-4164-5414]{Yonghwi Kim}
\altaffiliation{Jeonghyeon Ahn and Yonghwi Kim contributed equally to this work as first authors.}
\affiliation{Korea Institute of Science and Technology Information, 245 Daehak-ro, Yuseong-gu, Daejeon 34141, Republic of Korea}

\email{postgauss@snu.ac.kr, unitree@snu.ac.kr,
yonghwi.kim@kisti.re.kr}

\begin{acronym}
  \acro{FFT}{fast Fourier transform}
  \acro{MHD}{magnetohydrodynamic}
  \acro{DGF}{discrete Green's function}
  \acro{CCGF}{compact cylindrical Green’s function}
  \acro{MPI}{Message Passing Interface}
\end{acronym}

\begin{abstract}
We present an efficient and accurate algorithm for solving the Poisson equation in spherical polar coordinates with a logarithmic radial grid and open boundary conditions. The method employs a divide-and-conquer strategy, decomposing the computational domain into hierarchical units with varying cell sizes. James’s algorithm is used to compute the zero-boundary potentials of lower-level units, which are systematically integrated to reconstruct the zero-boundary potential over the entire domain. These calculations are performed efficiently via matrix-vector operations using various precomputed kernel matrices. The open-boundary potential is then obtained by applying a discrete Green’s function to the effective screening density induced at the domain boundaries. The overall algorithm achieves a computational complexity of $\mathcal{O}(N^3 \log N)$, where $N$ denotes the number of cells in one dimension. We implement the solver in the \texttt{FARGO3D} magnetohydrodynamics code and demonstrate its performance and second-order accuracy through a series of test problems.
\end{abstract}


\section{Introduction} \label{sec:intro}

Self-gravity plays a crucial role in the evolution of various astronomical systems, including large-scale cosmological simulations \citep[e.g.,][]{HR5, SIMBA}, galaxy-galaxy interactions \citep[e.g.,][]{GM1, GM2}, gravitational instabilities and star formation in galactic disks \citep[e.g.,][]{kim02,kim06,GI,kim20}, and dynamics in protoplanetary disks \citep[e.g.,][]{DI1, DI2}. The mutual gravitational attraction between mass elements is fundamental in shaping the structure, dynamics, and evolution of these systems, affecting processes such as galaxy formation, disk instabilities, planet formation, etc. Therefore, realistic and reliable simulations require accurate and efficient calculation of self-gravity.

In grid-based simulations,  self-gravity is  calculated by solving the Poisson equation
\begin{equation}\label{eq:Poisson} 
\boldsymbol{\nabla}^{2}\Phi = 4\pi G \rho, 
\end{equation}
after discretizing the domain using finite differences. Here, $\Phi$ and $\rho$ represent the open-boundary gravitational potential and matter density, respectively.
The choice of discretization depends on grid geometry, with different operators used for structured and unstructured grids. These operators affect the accuracy and efficiency of the solution, while grid geometry also influences boundary condition complexity and computational cost. Therefore, solvers must be tailored to the grid structure for optimal performance.

Cartesian, cylindrical, and spherical polar coordinate systems are commonly used in astronomical simulations.  Cartesian grids are suitable for systems where rotation is negligible, while cylindrical grids are preferred when accurate angular momentum conservation is essential.   Spherical polar coordinates are advantageous for systems with strong spherical symmetry, such as stellar structures \citep[e.g.,][]{Envel} and supernova explosions \citep[e.g.,][]{SN}.  For modeling gravitational instabilities in flared protoplanetary disks -- where disk thickness increases with radius -- cylindrical grids suffer from poor vertical resolution near the center. In such cases, spherical polar grids with a limited polar angle range are more effective in resolving the physical features of the system.

Various methods have been developed to solve the Poisson equation across different coordinate systems. Typical solvers include components for handling both internal and boundary potentials. Among the most widely used approaches for computing internal potentials are the fast Fourier transform (FFT; \citet{FFT1}; \citet{FFT2}) and the multigrid method \citep{Multi}. The multigrid method, employed in codes such as $\texttt{Athena++}$ \citep{Athena}, $\texttt{CASTRO}$ \citep{CASTRO}, and $\texttt{NIRVANA}$ \citep{NIRVANA}, is known for its versatility and scalability. These solvers are typically implemented in Cartesian coordinates, and extending them to spherical polar coordinates is nontrivial. In addition, since the multigrid method iteratively reduces errors, its convergence speed depends on multiple factors. Proper boundary conditions are also essential for ensuring solver performance.

The FFT method simplifies the discrete Poisson equation by applying a Fourier transform along symmetric axes, enabling an exact solution under periodic boundary conditions. This approach is highly efficient and often outperforms multigrid methods in speed \citep[see, e.g.,][]{FFTvsMulti}. However, its applicability depends on grid geometry. In uniform Cartesian coordinates, Fourier transforms can be applied along all axes, reducing the Poisson equation to a simple linear relation between potential and density. In contrast, cylindrical coordinates lack symmetry in the radial direction, preventing a Fourier transform along that axis. As a result, the Poisson equation becomes a tridiagonal system \citep{Moon}, which can be efficiently solved using the Thomas algorithm.

In spherical polar coordinates, the Fourier transform is applicable only in the azimuthal direction, and the Poisson equation remains challenging to solve directly. \citet{Muller19} proposed an alternative approach that reformulates the equation into a tridiagonal system via a kernel matrix transformation, eliminating the need for an FFT along the polar axis. This method has a computational complexity of $\mathcal{O}(N^4)$, where $N$ is the grid size in one dimension, primarily due to matrix multiplications. Despite the high complexity, modern hardware accelerates matrix operations, making the method efficient for moderate grid sizes. However, the algorithm assumes full coverage of the polar angle ($0\leq \theta\leq \pi$) to simplify boundary potential evaluation, which becomes inefficient for systems with flared density distributions such as protoplanetary disks.

When the simulation domain spans only a limited range in the polar direction, alternative methods are needed to impose Dirichlet boundary conditions on a spherical polar grid. Two such methods are the multipole expansion \citep{MULTIPOLE1, MULTIPOLE2} and the compact cylindrical Green’s function (CCGF) method \citep{CCGF}. The multipole expansion approach, similar to the Poisson solver by \citet{Muller95}, expresses the Green’s function as a series of eigenfunctions truncated at a chosen multipole order. However, this method becomes computationally expensive when significant density is present near domain boundaries \citep{CCGF}. The CCGF method was developed to mitigate this issue, yet it also requires truncation at a specific multipole number, introducing ambiguity in selecting an optimal cutoff. Furthermore, in cases of strong non-axisymmetry, a high maximum multipole order may be needed \citep{From05}, increasing computational cost and memory usage.

The algorithm developed by \citet{James} provides an efficient method for computing boundary potentials and imposing Dirichlet boundary conditions. Conceptually analogous to the method of image charges in electromagnetics, the approach begins by solving \cref{eq:Poisson} for the zero-boundary potential $[\Phi]$, which is set to zero at the domain boundaries.\footnote{In this paper, the bracket symbol `$[\;]$' is used exclusively to denote the zero-boundary potential. Although conventionally used for definite integrals or commutators, it is adopted here for notational convenience.} From $[\Phi]$, the screening surface density $\varrho$ is computed at the boundaries via $\varrho = -\boldsymbol{\nabla}^{2}[\Phi]/4\pi G$. The open-boundary potential $\Theta$, generated by $\varrho$, contributes to the total potential through the relation $\Phi$ through the relation $\Phi=[\Phi]+\Theta$. Notably, $\Theta$ directly represents the potential at the boundaries where $[\Phi]=0$, allowing it to be evaluated by applying the Green’s function to $\varrho$, which then serves as the Dirichlet boundary condition. \citet{Moon} developed an efficient method to compute gravitational potentials in cylindrical coordinates by using FFT to calculate $[\Phi]$ and combining discrete Green’s functions (DGF) with FFT to compute $\Theta$. More recently, \citet{Ziegler24} extended this approach to spherical polar coordinates, employing a multigrid solver to obtain $[\Phi]$. 

In multi-threaded simulations, parallel communication is just as critical as the computational workload. Parallelized FFT operations involve substantial inter-core communication, and minimizing this overhead can significantly enhance performance. For instance, in simulations of flared astrophysical disks using \texttt{FARGO3D}, a high-performance MHD code operating in cylindrical or spherical polar coordinates \citep{FARGO1,FARGO2}, efficiency is improved by avoiding grid partitioning in the azimuthal direction. This design eliminates inter-core communication during azimuthal FFTs, enabling efficient execution. Under this partitioning scheme, FFT-based methods in the azimuthal direction outperform alternative approaches.

In this paper, we present a novel Poisson solver for \texttt{FARGO3D} in spherical polar geometry. The algorithm employs James’s method to impose Dirichlet boundary conditions and compute internal gravitational potentials across the entire domain. We adopt a divide-and-conquer strategy by partitioning the computational domain into hierarchical units with varying cell sizes. The zero-boundary potential is constructed by recursively integrating contributions from lower-level units. The merging process involves a series of linear transformations using precomputed kernel matrices to enhance performance. The overall computational complexity of the method is $\mathcal{O}(N^3 \log N)$, and it significantly outperforms MHD calculations in terms of runtime. Parallelization is implemented using the Message Passing Interface (MPI), following \texttt{FARGO3D}’s predefined grid partitioning scheme. Kernel data are shared via shared memory within each node to reduce redundancy. Since the solver evaluates density and potential only at the boundaries at each hierarchical level, the inter-core communication overhead is lower than that of MHD solvers. We validate the solver's efficiency and demonstrate second-order accuracy through a series of benchmark tests. 

This paper is organized as follows. \autoref{sec:equation} presents the discretized, Fourier-transformed Poisson equation in spherical polar coordinates that we solve. \autoref{sec:method} describes our divide-and-conquer strategy for calculating the open-boundary potential and the screening density induced at the domain boundaries. \autoref{sec:DGF} explains the DGF necessary for calculating the open-boundary potential from the screening density. \autoref{sec:OverannImple} provides an overview of the steps in our Poisson solver and its implementation in \texttt{FARGO3D}. \autoref{sec:tests} presents the results of our Poisson solver on various test problems to demonstrate its accuracy and efficiency. Finally, \autoref{sec:summary} summarizes and discusses the present work.

\section{Reduced Equation} \label{sec:equation}

Our computational domain is a sector in spherical polar coordinates $(r, \theta, \phi)$, bounded by $[r_\text{min}, r_\text{max}]$, $[\theta_\text{min},\theta_\text{max}]$, and $[0, 2\pi ]$ in the radial, polar, and azimuthal directions, respectively. We enforce the condition $\theta_\text{min}+\theta_\text{max}=\pi$ to ensure symmetry about the equatorial plane. We discretize the domain into $N_r\times N_\theta \times N_\phi$ cells. In the radial direction, we set up a logarithmically spaced grid with face centers at $r_{i+1/2}=f^{i} r_\text{min}$ for $i = 0, 1, 2, \dots, N_r-1$, where $f \equiv (r_\text{max}/r_\text{min})^{1/N_r} > 1$ is the multiplication factor. The cell-centered radial coordinates are defined using volumetric centers as 
\begin{equation}
r_{i} = \frac{\int^{r_{i+1/2}}_{r_{i-1/2}} 
r^3 dr}{\int^{r_{i+1/2}}_{r_{i-1/2}} 
r^2 dr}
= \frac{3(f^2+1)(f+1)}{4(f^2+f+1)}f^{i-1} r_{\text{min}}.
\end{equation}
In the polar direction, we define a uniformly spaced grid with face centers at $\theta_{j+1/2}=\theta_\text{min} + j \delta \theta$ for $j = 0, 1, 2, \dots, N_{\theta}-1$ and cell centers at 
$\theta_{j} = (\theta_{j- 1/2} +\theta_{j+ 1/2})/2$, where 
$\delta \theta = (\theta_\text{max}-\theta_\text{min})/N_\theta$. 
In the azimuthal direction, we use a uniformly spaced grid with face centers at  $\phi_{k+1/2}=  k \delta \phi$ for $k = 0, 1, 2, \dots, N_{\phi}-1$ and cell centers at $\phi_{k}=(\phi_{k- 1/2} + \phi_{k+ 1/2})/2$, where $\delta \phi= 2\pi/N_\phi$.

The second-order finite-difference approximation to \cref{eq:Poisson} is given by \begin{equation}\label{eq:dPos}
\left(\Delta^{2}_{r} + 
\Delta^{2}_{\theta} + 
\Delta^{2}_{\phi}\right)\Phi_{i,j,k} = 4\pi G \rho_{i,j,k},
\end{equation}
where $\Delta^{2}_{r}$, $\Delta^{2}_{\theta}$, and $\Delta^{2}_{\phi}$ refer to the second-order Laplace operator in each coordinate direction, given by
\begin{equation}\label{eq:genFDr}
\begin{split}
\Delta^{2}_{r}\Phi_{i,j,k} &= \frac{3}{r^{3}_{i+1/2}-r^{3}_{i-1/2}} \left(\frac{\Phi_{i+1,j,k}-\Phi_{i,j,k}}{r_{i+1}-r_{i}}r^{2}_{i+1/2} - \frac{\Phi_{i,j,k}-\Phi_{i-1,j,k}}{r_{i}-r_{i-1}}r^{2}_{i-1/2}\right) \\
 &= 
\frac{2f \,\mathcal{R}_i}{(f^2-1)(f-1)}
\left[
f\Phi_{i+1,j,k}-(f+1)\Phi_{i,j,k}+\Phi_{i-1,j,k}
\right],
\end{split}
\end{equation}
\begin{equation}\label{eq:genFDt}
\Delta^{2}_{\theta}\Phi_{i,j,k} = \frac{\mathcal{R}_i}{(\cos\theta_{j-1/2}-\cos\theta_{j+1/2})} \bigg( \frac{\Phi_{i,j+1,k}-\Phi_{i,j,k}}{\theta_{j+1}-\theta_{j}}\sin\theta_{j+1/2}
- \frac{\Phi_{i,j,k}-\Phi_{i,j-1,k}}{\theta_{j}-\theta_{j-1}}\sin\theta_{j-1/2} \bigg),
\end{equation}
\begin{equation}\label{eq:genFDp}
\Delta^{2}_{\phi}\Phi_{i,j,k} = \frac{\mathcal{R}_i \mathcal{S}_j}{\delta \phi} \bigg( \frac{\Phi_{i,j,k+1}-\Phi_{i,j,k}}{\phi_{k+1}-\phi_{k}}
- \frac{\Phi_{i,j,k}-\Phi_{i,j,k-1}}{\phi_{k}-\phi_{k-1}} \bigg),
\end{equation}
with
\begin{equation}\label{eq:genR}
\mathcal{R}_i \equiv \frac{3(r^{2}_{i+1/2} - r^{2}_{i-1/2})}{2r_{i}(r^{3}_{i+1/2}-r^{3}_{i-1/2})}
= \frac{2f^{2-2i}}{(f^2+1)r_\text{min}^2},
\end{equation} 
and 
\begin{equation}
\mathcal{S}_j \equiv \frac{\delta\theta}{\sin\theta_{j}(\cos\theta_{j-1/2}-\cos\theta_{j+1/2})}
\end{equation}
\citep[e.g.,][]{Muller19}. Note that the radial dependence of all three Laplace operators can be handled simply by $\mathcal{R}_i$, which depends only on the index $i$, when the radial grid is logarithmically spaced. This helps reduce memory usage for the various kernels discussed in \autoref{sec:method}.

We apply a Fourier transform in the azimuthal direction to reduce the three-dimensional \cref{eq:dPos} to a two-dimensional (2D) equation.
By imposing periodic boundary conditions in the azimuthal direction, we decompose $\Phi_{i,j,k}$ and $\rho_{i,j,k}$ as 
\begin{equation}\label{eq:FFT}
\Phi_{i,j,k}= \sum_{m=1}^{N_\phi} \Phi_{i,j}^m \mathcal{P}^{m}_{k},
\qquad \text{and} \qquad 
\rho_{i,j,k}= \sum_{m=1}^{N_\phi} \rho_{i,j}^m \mathcal{P}^{m}_{k},
\end{equation}
where 
$\Phi_{i,j}^m$ and $\rho_{i,j}^m$ are the Fourier transform of $\Phi_{i,j,k}$ and $\rho_{i,j,k}$, respectively, and 
\begin{equation}
\mathcal{P}^{m}_{k} = \exp \left[ \frac{2\pi \sqrt{-1}mk}{N_{\phi}} \right]
\end{equation}
is the eigenfuntion for the discrete Laplace operator $\Delta_\phi^2$. In terms of $\Phi_{i,j}^m$ and $\rho_{i,j}^m$, \cref{eq:dPos} reduces to 
\begin{equation}\label{eq:DPE}
\left(\Delta^{2}_{r} + 
\Delta^{2}_{\theta} + 
\lambda^{m}_{i,j}\right)\Phi^{m}_{i,j} = 4\pi G \rho^{m}_{i,j},
\end{equation}
where 
\begin{equation}\label{eq:lambdam}
\lambda^{m}_{i,j} = - \mathcal{R}_i \mathcal{S}_j \left[ \sin\left(\frac{m\pi}{N_{\phi}}\right) \Big/ \frac{\pi}{N_{\phi}} \right]^{2}
\end{equation} 
is the eigenvalue for the operator $\Delta_\phi^2$. \Cref{eq:DPE} is the reduced form of \cref{eq:dPos}, representing $N_{\phi}$ independent equations on a 2D grid spanned by the radial and polar directions.

\section{Divide-and-conquer Method}\label{sec:method}

For each azimuthal mode number $m$, \cref{eq:DPE} represents a Poisson equation for the Fourier-transformed potential $\Phi_{i,j}^m$ induced by the Fourier-transformed mass distribution $\rho_{i,j}^m$ in a 2D domain. In this section, we develop a divide-and-conquer method to solve \cref{eq:DPE} efficiently, using hierarchical units with varying cell sizes. For simplicity, we omit the subscripts `$i$' and `$j$' and superscript `$m$' in this section.

\begin{deluxetable}{@{\extracolsep{4pt}}cccccc}
\tablecaption{Size of units at each level for  $n_r=\log_2 (N_r+1)$ and $n_\theta = \log_2 (N_\theta+1)$.
\label{tab:level}}
\tablehead{
\multicolumn{2}{c}{$n_r = n_\theta$} &
\multicolumn{2}{c}{$n_r > n_\theta$} &
\multicolumn{2}{c}{$n_r < n_\theta$} \\
\cline{1-2} \cline{3-4} \cline{5-6}
\colhead{Level} & \colhead{unit size} &
\colhead{Level} & \colhead{unit size} &
\colhead{Level} & \colhead{unit size} 
}
\startdata
0 & $3 \times 3$ &
0 & $3 \times 3$ & 
0 & $3 \times 3$\\
1 & $3 \times 7$ &
1 & $3 \times 7$ & 
1 & $3 \times 7$\\
2 & $7 \times 7$ &
2 & $7 \times 7$ & 
2 & $7 \times 7$\\
\vdots & \vdots & \vdots &
\vdots & \vdots & \vdots \\
$2k$ & $(2^{k+2}-1) \times (2^{k+2}-1)$ &
$2n_\theta - 4$ & $(2^{n_\theta}-1) \times 2^{n_\theta}$ &
$2n_r - 4$ & $(2^{n_r}-1) \times (2^{n_r}-1)$ \\
$2k+1$ & $(2^{k+2}-1) \times (2^{k+3}-1)$ &
$2n_\theta - 3$ & $(2^{n_\theta+1}-1) \times 2^{n_\theta}$ &
$2n_r - 3$ & $(2^{n_r}-1) \times (2^{n_r+1}-1)$ \\
\vdots & \vdots & \vdots &
\vdots & \vdots & \vdots \\
$n_\theta + n_r - 5$ & $(2^{n_r-1}-1) \times 2^{n_\theta}$ &
$n_\theta + n_r - 5$ & $(2^{n_r-1}-1) \times 2^{n_\theta}$ &
$n_\theta + n_r - 5$ & $(2^{n_r}-1) \times (2^{n_\theta-1}-1)$\\
$n_\theta + n_r - 4$ & $(2^{n_r}-1) \times 2^{n_\theta}$ &
$n_\theta + n_r - 4$ & $(2^{n_r}-1) \times 2^{n_\theta}$ &
$n_\theta + n_r - 4$ & $(2^{n_r}-1) \times 2^{n_\theta}$\\
\enddata
\end{deluxetable}

\subsection{Overview of James's Algorithm}

We first briefly review James's algorithm, which forms the foundation of our Poisson solvers for calculating the potential of units at various levels. For any 2D unit with $I\times J$ cells, James's method can be applied to compute the open-boundary solution $\Phi$ of \cref{eq:DPE}. The screening densities $\varrho_T$, $\varrho_B$, $\varrho_L$, and $\varrho_R$ in the top, bottom, left, and right boundaries can be calculated as 
\begin{align}\label{eq:scrdens2D}
  \begin{aligned}
   \varrho_T &= -\frac{\mathcal{R}_i \sin \theta_{j+1/2}}{(\cos\theta_{j+1/2}-\cos\theta_{j+3/2}) \delta \theta}[\Phi]_{j=J}, \\    
   \varrho_L &= -\frac{3r^{2}_{i-1/2}}{(r^{3}_{i-1/2}-r^{3}_{i-3/2}) \delta r} [\Phi]_{i=1},
  \end{aligned}
  &&
  \begin{aligned}
   \varrho_B &= -\frac{\mathcal{R}_i \sin \theta_{j-1/2}}{(\cos\theta_{j-3/2}-\cos\theta_{j-1/2}) \delta \theta}[\Phi]_{j=1}, \\       
   \varrho_R &= -\frac{3r^{2}_{i+1/2}}{(r^{3}_{i+3/2}-r^{3}_{i+1/2}) \delta r} [\Phi]_{i=I},
  \end{aligned}
 \end{align}
respectively. The open-boundary potential $\Phi$ is then given by 
\begin{equation}\label{eq:james}
 \Phi = [\Phi] + \Theta_{T} + \Theta_{B} + \Theta_{L} + \Theta_{R},
\end{equation}
where $\Theta_{T}$, $\Theta_{B}$, $\Theta_{L}$, and $\Theta_{R}$ are the potentials generated by the screening densities $\varrho_{T}$, $\varrho_{B}$, $\varrho_{L}$, and $\varrho_{R}$, respectively. We will use \cref{eq:james} to compute the potentials of various units described below. 

\subsection{Grid Partitioning and Level Structure}
 
In our divide-and-conquer strategy, we begin by partitioning the entire 2D domain into its smallest units at the lowest level\footnote{Here, 'level' refers to the recursive depth in the divide-and-conquer hierarchy, and should not be confused with grid resolution levels used in AMR or multigrid methods. In an AMR grid structure, for example, the coarsest (largest) grid is typically defined as Level 0, whereas in the algorithm presented in this paper, Level 0 corresponds to the finest (smallest) partition.}, computing the potentials within these units and along their boundaries. We then iteratively merge the potentials from lower levels to determine the potentials at higher levels, continuing this process until the full potential across the entire grid is obtained.

\begin{figure} 
	\centering
    \includegraphics[width=13cm]{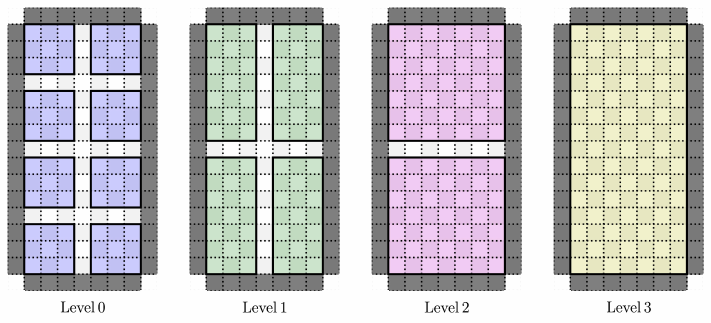}
    \caption{Illustration of our grid partitioning and level structure for the case of a $7\times 15$ grid. Connected colored cells form units at each level, while white cells represent intervening boundaries at the same level. The boundaries of the entire domain are shaded in gray.
    \label{fig:level}} 
\end{figure}

The smallest units in our strategy consist of $3\times3$ cells with intervening boundaries, as illustrated in \cref{fig:level}  when the entire grid contains $7\times 15$ cells. At Level 0, the computational domain is divided into eight smallest units with their boundaries. Two neighboring smallest units, together with their intervening boundary cells, are merged vertically to form four $3\times 7$ units, which we refer to as the grid structure at Level 1. At Level 1, two neighboring $3\times 7$ units, including their intervening boundary, are merged horizontally to form two $7\times7$ units, referred to as Level 2. Finally, two $7\times7$ units with their intervening boundary are merged vertically to form the complete grid at Level 3.  

Our level and grid structure requires $N_r \times  N_\theta = (2^{n_r} -1)\times (2^{n_\theta}-1)$ for integers $n_r,n_\theta\geq 2$.\footnote{See \autoref{sec:power} for the case with $N_r \times  N_\theta = 2^{n_r}\times 2^{n_\theta}$.} 
\autoref{tab:level} presents the unit sizes at each level, depending on the relative magnitudes of $n_r$ and $n_\theta$. When units are available for merging in both directions, the merging direction alternates with increasing level number. If no further units remain in one direction, which occurs at higher levels when $n_r\neq n_\theta$, merging continues in the other direction. In all cases, the highest level is Level $n_r+n_\theta-4$, and the number of unit mergers in the radial and polar directions are $n_r-2$ and $n_\theta-2$, respectively. \Cref{fig:level} corresponds to the case with $n_r=3$ and $n_\theta=4$.

\subsection{Unit Merging}

\begin{figure} 
\centering
\includegraphics[width=7.0cm]{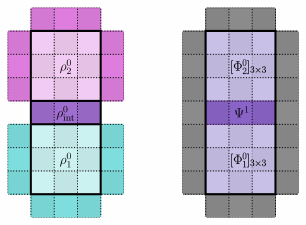}
    \caption{Illustration of the merging of two units at Level 0 (left) into a single unit at Level 1 (right). At Level 0, $\rho_1^0$ and $\rho_2^0$ represent the density of the lower and upper units, with the corresponding zero-boundary potentials $[\Phi_1^0]_{3\times 3}$ and $[\Phi_2^0]_{3\times 3}$, respectively. Note that $\rho_\text{int}^0$ denotes the density in the intervening boundary at Level 0, which together with the screening charges $\rho_{1,T}$ and $\rho_{2,B}$ in the middle line at Level 1 produces the zero-boundary potential $\Psi^1$.
    \label{fig:level1}} 
\end{figure}

Suppose we have two neighboring units at Level 0, as illustrated in the left panel of \cref{fig:level1}. \Cref{eq:james} implies that the zero-boundary potential in the lower unit with size $3\times3$ cells is given by 
\begin{align}\label{eq:merge33}
[\Phi^0_1]_{3\times 3} =    \Phi^0_1 - \Theta^0_{1,T} - \Theta^0_{1,B} - \Theta^0_{1,L} - \Theta^0_{1,R} \equiv \Phi^0_{1,\text{eff}},
\end{align}
where the superscript `0' denotes the level number and the subscript `1' indicates the lower unit. Note that zero-boundary potential $[\Phi^0_1]_{3\times 3}$ is non-zero only within the 9-cell lower unit and vanishes elsewhere, and that the open-boundary potentials induced by the screening densities are not restricted to the lower unit. \Cref{eq:merge33} indicates that the zero-boundary potential due to $\rho_1^0$ within the unit is equivalent to the open-boundary potential $\Phi^0_{1,\text{eff}}$ due to the effective density distribution, defined as $\rho_{1,\text{eff}}^0 \equiv \rho_1^0 - \varrho^0_{1,T} - \varrho^0_{1,B} - \varrho^0_{1,L} - \varrho^0_{1,R}$. This effective density is distributed over the $5\times 5$ region, consisting of the lower unit and its surrounding boundaries.

Since $\Phi_{1,\text{eff}}^0=0$ outside the lower unit, \cref{eq:merge33} leads to an important result
\begin{equation}\label{eq:merge33b}
[\Phi_1^0]_{3 \times 3}  =  [\Phi^0_{1,\text{eff}}]_{3 \times 7}, 
\end{equation}
where the $3 \times 7$ region encompasses the lower and upper units along with the intervening boundary. Imposing the zero-boundary condition on \cref{eq:merge33} within the $3 \times 7$ region and utilizing \cref{eq:merge33b} yields
\begin{equation}\label{eq:merge33c}
   [\Phi^0_{1}]_{3 \times 7} = [\Phi_1^0]_{3 \times 3} + [\Theta^0_{1,T}]_{3 \times 7},
\end{equation}
since only the top boundary among the four boundaries is included in the $3 \times 7$ region. For the upper unit, an analogous relation holds:
\begin{equation}\label{eq:merge33d}
   [\Phi^0_{2}]_{3 \times 7} = [\Phi_2^0]_{3 \times 3} + [\Theta^0_{2,B}]_{3 \times 7},
\end{equation}
where $[\Phi_2^0]_{3 \times 3}$ represents the zero-boundary potential due to the density $\rho_2^0$ in the upper unit and  $[\Theta^0_{2,B}]_{3\times7}$ denotes the zero-boundary potential arising from the screening density in the bottom boundary of the upper unit at Level 0.

We now consider the same $3 \times 7$ region as a new unit at Level 1, with the density distribution given by $\rho^1 \equiv \rho^0_1 + \rho^0_2 + \rho^0_\text{int}$ (see the right panel of \cref{fig:level1}). Here, $\rho^0_\text{int}$ represents the matter density at the interface between the lower and upper units at Level 0. Since the Poisson equation is linear, the corresponding zero-boundary potential is given by 
\begin{equation}\label{eq:7times3}
[\Phi^1]_{3 \times 7} = [\Phi^0_1]_{3 \times 7} + [\Phi^0_2]_{3 \times 7} + [\Phi^0_\text{int}]_{3 \times 7} 
= [\Phi^0_1]_{3 \times 3} + [\Phi^0_2]_{3 \times 3} + \Psi^1,
\end{equation}
where $[\Phi^0_\text{int}]_{3 \times 7}$ 
is the zero-boundary potential due to $\rho^0_\text{int}$ and  
\begin{equation}
\Psi^1 \equiv [\Theta^0_{1,T}]_{3 \times 7} + [\Theta^0_{2,B}]_{3 \times 7} + [\Phi^0_\text{int}]_{3 \times 7}
\end{equation}
is the zero-boundary potential generated by the effective density 
\begin{equation}\label{eq:rholine}
\rho^1_\text{mid} \equiv  \varrho^0_{1,T} + \varrho^0_{2,B} + \rho^0_\text{int}
\end{equation} 
in the middle line. \cref{eq:7times3} states that the zero-boundary potential of a unit at a given level is the sum of the zero-boundary potentials of its two subordinate units at the lower level and the zero-boundary potential induced by the effective density at the middle-line cells.

\begin{figure} 
	\centering
    \includegraphics[width=13cm]{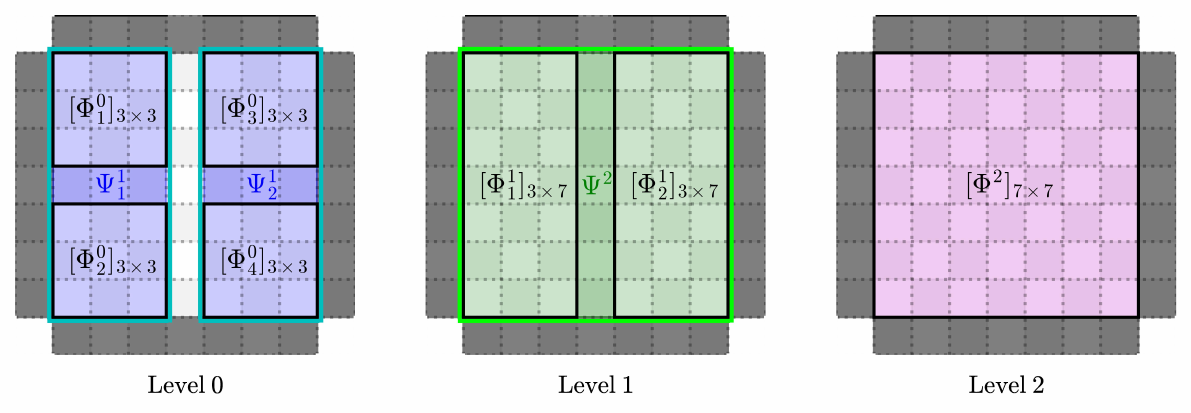}
    \caption{Illustration of the merging process, where four units at Level 0, each with zero-boundary potentials $[\Phi^0_i]_{3\times3}$ and line-density potentials $\Psi^1_i$ (left), combine into two units at Level 1 with zero-boundary potentials $[\Phi^1_i]_{3\times7}$ (middle). These units are further merged, incorporating $\Psi^2$, into a single unit at Level 2 possessing the zero-boundary potential $[\Phi^2_i]_{7\times7}$ (right).
    \label{fig:level2}} 
\end{figure}

By the same logic, the zero-boundary potential of a unit at Level 2, shown in \cref{fig:level2}, is given by 
\begin{equation}\label{eq:7times7_1}
[\Phi^2]_{7\times 7} = [\Phi^1_1]_{3\times 7}
+ [\Phi^1_2]_{3\times 7} + \Psi^2,
\end{equation} 
where the subscripts `1' and `2' denote the left and right units at Level 1, respectively, and $\Psi^2$ is the zero-boundary potential due to the effective density in the middle line at Level 2.  Using \cref{eq:7times3} for the left and right units at Level 1, \cref{eq:7times7_1} can be written as 
\begin{equation}\label{eq:7times7}
[\Phi^2]_{7\times 7} = \sum_{i=1}^4 [\Phi^0_i]_{3\times 3} + \Psi^1_1 + \Psi^1_2 + \Psi^2,
\end{equation}
where $[\Phi^0_i]_{3\times 3}$ indicates the zero-boundary potential of four units at Level 0, while $\Psi^1_1$ and $\Psi^1_2$ denote the zero-boundary potentials arising from the effective line density along the middle line of the left and right units at Level 1, respectively.

Generalizing this procedure, the zero-boundary potential of a unit at Level $k$ can be expressed as  
\begin{equation}\label{eq:mergepot}
[\Phi^k] = \sum_i^{2^k} [\Phi^0_i] + \sum_{l=1}^{k} \sum_{i=1}^{2^{k-l}} \Psi^l_i,
\end{equation}
where unit sizes for the zero-boundary potentials are omitted for simplicity. \Cref{eq:mergepot} shows that $[\Phi^k]$ is obtained by summing the zero-boundary potentials $[\Phi^0_i]$ of all units at Level 0 and the potentials $\Psi^l_i$ associated with middle-line densities across Level $k$ and lower. In what follows, we describe the computation of $[\Phi^0_i]$ and $\Psi^l_i$.

\subsection{Potentials of Smallest Units}\label{sec:PSU}

To calculate the zero-boundary potential of the smallest units, we introduce a density-to-potential (D2P) kernel matrix.  Under the zero-boundary condition,  \cref{eq:DPE} applied to a unit consisting of $3\times3$ cells at Level 0 can be formally expressed as 
\begin{equation}\label{eq:matrix1}
\mathcal{A}_\text{D2P} \rho^0 = [\Phi^0],
\end{equation}
where ${\mathcal A}_\text{D2P}$ is the $9\times9$ D2P kernel matrix, and $\rho^0$ and $[\Phi]$ are column vectors of length nine, representing the density and zero-boundary potential of the corresponding cells in the unit. These kernel matrices for different locations and Fourier mode numbers $m$ are precomputed once and stored before the main simulation loop begins.

For given $m$, there are $2^{n_r+n_\theta-4}$ units at Level 0. Since the Laplace operators depend on $r$, $\theta$, and $\phi$, the kernel matrices generally vary across units. However, if the grid retains equatorial symmetry, the same kernel matrices can be used for both the northern and southern hemispheres. Additionally, if the radial grid is logarithmically spaced, the radial dependence of the kernels can be accounted for by the scale factor $\mathcal{R}_i$ defined in \cref{eq:genR}. By utilizing this symmetry, the total number of D2P kernel matrices that must be stored at Level 0 is $N_\phi 2^{n_\theta-3}$.
The number of operations required to compute $[\Phi_i^0]$ from $\rho_i^0$ is of order of ${\mathcal O}(N_r N_\phi N_\theta)$.

To determine the elements of ${\mathcal A}_\text{D2P}$, we consider a case where the density is confined to a single cell, say, $\rho_1^0=1$ while $\rho_{i}^0=0$ for $i=2,3,\cdots,9$. We calculate $[\Phi^0]$ resulting from this single-cell density using the projection method of \citet{Muller19}, as outlined in \autoref{sec:proj}. Then, \cref{eq:matrix1} provides the first column of ${\mathcal A}_\text{D2P}$. By systematically shifting the unit density across all cells in the unit, we determine all 81 elements of ${\mathcal A}_\text{D2P}$. Once all the kernel matrices are established, computing $[\Phi_i^0]$ from $\rho_i^0$ becomes straightforward.

\subsection{Potentials Due to Line Densities}
\label{sec:Zerodens}

In the second term on the right-hand side of  \cref{eq:mergepot}, $\Psi^l$ represents the zero-boundary potential generated by the effective densities along the middle line of a unit at Level $l$. We compute $\Psi^l$ efficiently using a divide-and-conquer strategy, as detailed below.

\begin{figure} 
	\centering
    \includegraphics[width=13cm]{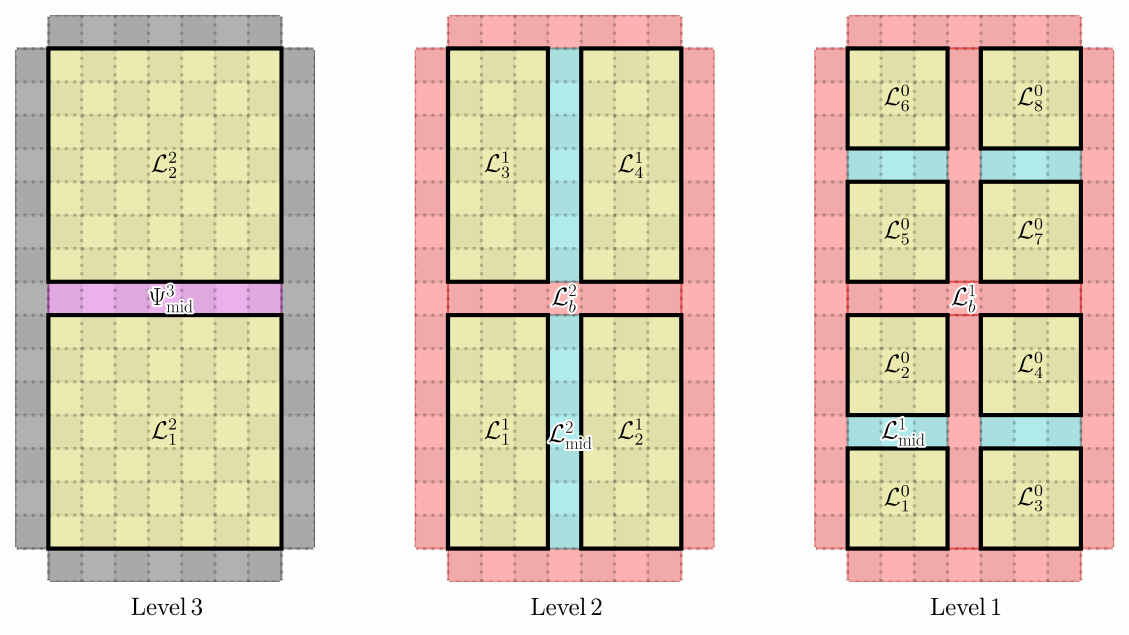}
    \caption{The regions shaded in yellow and red correspond to the areas used for calculating the source-free interior potentials, $\mathcal{L}_i^{l-1}$, and source-free boundary potentials, $\mathcal{L}_b^l$, respectively. The purple-highlighted region denotes the area used to compute the middle-line potential, $\Psi_\text{mid}^3$.
    \label{fig:lap}}
\end{figure}

\Cref{fig:lap} illustrates the procedure for computing $\Psi^l$ when $l=3$. Note that the effective density for $\Psi^l$, denoted as $\rho_\text{mid}^l$, is non-zero along the middle line and zero elsewhere (see \autoref{eq:rholine}). We split $\Psi^l$ into three terms as 
\begin{equation}\label{eq:Psil}
\Psi^l = \mathcal{L}^{l-1}_1 + \mathcal{L}^{l-1}_2 + \Psi^l_\text{mid},
\end{equation}
where $\mathcal{L}^{l-1}_1$ and $\mathcal{L}^{l-1}_2$ represent the potentials in the two subordinate units at Level $l-1$, and $\Psi^l_\text{mid}$ denotes the potential along the middle line. We calculate $\Psi^l_\text{mid}$ using 
\begin{equation}\label{eq:d2p}
{\mathcal K}^l_\text{D2P} \rho^l_\text{mid} = \Psi^l_\text{mid},
\end{equation}
where ${\mathcal K}^l_\text{D2P}$ is another D2P kernel matrix at Level $l$. This square matrix has a number of rows equal to the number of cells in the middle line. Similarly to ${\mathcal A}_\text{D2P}$ in \cref{eq:matrix1}, ${\mathcal K}^l_\text{D2P}$ can be precomputed using single-cell densities along the middle line, with the potential obtained via the projection method of \citet{Muller19}. The number of ${\mathcal K}^l_\text{D2P}$ matrices that need to be stored at Level $l$ is $N_\phi N_\theta / 2^{(l+7)/2}$ for odd $l$ and $N_\phi N_\theta / 2^{l/2+3}$ for even $l$.

Since the internal density is zero except along the middle line at Level $l$, the potentials $\mathcal{L}^{l-1}_1$ and $\mathcal{L}^{l-1}_2$ must satisfy the Laplace equation, meaning their Laplacian is zero. We further decompose each of these potentials into three parts.  For $\mathcal{L}_1^{l-1}$, we write 
\begin{equation}\label{eq:LL}
\mathcal{L}^{l-1}_1 = \mathcal{L}^{l-2}_{1} + \mathcal{L}^{l-2}_{2} + \mathcal{L}^{l-1}_\text{mid},
\end{equation}
where $\mathcal{L}^{l-2}_{1}$ and $\mathcal{L}^{l-2}_{2}$ represent the potentials in the two subordinate units at Level $l-2$, and $\mathcal{L}^{l-1}_\text{mid}$ denotes the potential along the middle line, as shown in the middle panel of \Cref{fig:lap}.  While \cref{eq:Psil,eq:LL} appear similar, we note that $\mathcal{L}_\text{mid}^{l-1}$ is source-free, whereas $\Psi_\text{mid}^l$ is not. 
Consequently, $\mathcal{L}_\text{mid}^{l-1}$ is entirely determined by the boundary potential $\mathcal{L}_b^{l-1}$ such that $\mathcal{L}_b^{l-1}=\Psi^l_\text{mid}$ along the middle line and $\mathcal{L}_b^{l-1}=0$ at the other three boundaries. We thus introduce a potential-to-potential (P2P) kernel matrix such that 
\begin{equation}\label{eq:p2p}
{\mathcal K}^{l}_\text{P2P} \mathcal{L}^{l}_b = \mathcal{L}^{l}_\text{mid}. 
\end{equation}
The calculation of the kernel matrices ${\mathcal K}^{l}_\text{P2P}$ is detailed in \autoref{sec:p2p}. 
If a unit at Level $l$ has dimensions $(2^p-1) \times (2^q-1)$, where $p$ and $q$ are integers and the middle-line length is $2^p-1$, then ${\mathcal K}^{l}_\text{P2P}$ contains $(2^p-1) \times (2^{p+1}+2^{q+1}-4)$ elements. The number of P2P kernels at Level $l$ is $N_\phi N_\theta /2^{q+1}$, which is equal to the number of D2P kernels.

We continue this division process until reaching Level 0.  Let $\mathcal{L}^0$ represent the interior potential within a unit consisting of $3\times3$ cells at Level 0. Since $\mathcal{L}^0$ is also a source-free interior potential, it can be determined directly from the boundary potentials $\mathcal{L}_b^0$ in the twelve surrounding cells along the four sides of the unit, given by
\begin{equation}\label{eq:p2p_0}
\mathcal{B}_\text{P2P} \mathcal{L}_b^0 = \mathcal{L}^0,
\end{equation}
where $\mathcal{B}_\text{P2P}$ is another P2P kernel matrix at Level 0, containing $9\times12$ elements. Analogous to ${\mathcal K}^{l}_\text{P2P}$, $\mathcal{B}_\text{P2P}$ can be computed using the single-cell potential across all twelve surrounding boundary cells.  Similarly to $\mathcal{A}_\text{D2P}$ in \cref{eq:matrix1}, the total number of $\mathcal{B}_\text{P2P}$ matrices that need to be stored across the entire grid is $N_\phi 2^{n_\theta-3}$.

Once all interior potentials $\mathcal{L}_i^0$ at Level 0 and middle-line potentials $\mathcal{L}_{\text{mid},i}^k$ at various levels are obtained, the zero-boundary potential $\Psi^l$ at Level $l$, induced by the effective density, can be constructed as  
\begin{equation}\label{eq:Psi_sum}
\Psi^l = \sum_{i=1}^{2^l} \mathcal{L}_i^0 
+ \sum_{k=1}^{l-1} \sum_{i=1}^{2^{l-k}}  \mathcal{L}_{\text{mid},i}^k + \Psi_\text{mid}^l,
\end{equation}
where the subscript `$i$' denotes different units at a given level. \cref{eq:mergepot,eq:Psi_sum} are combined to yield the zero-boundary potential at the highest level $L$ as
\begin{equation}\label{eq:Pot_psi}
[\Phi^L] =\sum_{i=1}^{2^L} [\Phi^0_i] 
+ \sum_{l=1}^{L} \sum_i \mathcal{L}^0_i(l) 
+ \sum_{l=1}^{L} \sum^{l-1}_{k=1} \sum_i \mathcal{L}^k_{\text{mid},i}(l)
+ \sum_{l=1}^{L} \sum_i \Psi^l_{\text{mid},i},
\end{equation}
where $\mathcal{L}^0(l)$  and $\mathcal{L}^k_{\text{mid}}(l)$ stand for the source-free boundary potential at Level 0 and the middle-line potential at Level $k$, generated during the division process in the computation of $\Psi^l$, respectively.

\subsection{Optimization}\label{sec:optim}

The method of calculating $\Psi^l$ at Level $l$, as described in the previous section, follows a top-down approach, suggesting that certain steps for $\Psi^l$ can be integrated with those for $\Psi^k$ at lower levels ($k<l$) to enhance computational efficiency. For instance, the left-hand side of \cref{eq:p2p} at Level  $1$ in the computation of $\Psi^{10}$ in \cref{eq:Psi_sum} can be combined with those for $\Psi^k$'s as 
\begin{equation}\label{eq:opt}
\sum_{k=1}^{10} \mathcal{K}_\text{P2P}^1 \mathcal{L}_b^1(k) 
= \mathcal{K}_\text{P2P}^1\sum_{k=1}^{10} 
 \mathcal{L}_b^1(k).
\end{equation}
Since matrix-vector multiplications are significantly more expensive than vector additions, this approach aims to drastically reduce the computational cost of evaluating the second and third terms in \cref{eq:Pot_psi}.

However, this approach presents a challenge, as the middle-line density $\rho_\text{mid}^l$ needed to find $\Psi_\text{mid}^l$ must be determined a priori before applying the D2P kernel in \cref{eq:d2p}. This requires obtaining $[\Phi^{l-1}]$ for all units at Level $l-1$, which in turn necessitates knowing $\Psi^{l-1}$ via \cref{eq:mergepot}, then $\Psi^{l-2}$, and so forth. One can naively calculate the zero-boundary potential by applying \cref{eq:mergepot,eq:Psi_sum} step by step from $l=0$ to the highest level. However, this approach fails to take advantage of the efficient computation discussed above. 

To this end, we develop a method to derive $\rho^l_\text{mid}$ without computing $[\Phi^{l-1}]$. Similarly to \cref{eq:rholine}, we express
\begin{equation}\label{eq:modirhoex}
\rho^l_{\text{mid}} = \varrho^{l-1}_1 + \varrho^{l-1}_2 + \rho^{l-1}_\text{int},
\end{equation}
where the first two terms represent the screening densities induced by the zero-boundary potentials  $[\Phi_1^{l-1}]$ and $[\Phi_2^{l-1}]$ in the two subordinate units at Level $l-1$, while the last term corresponds to the actual density along the middle line at Level $l$. Analogous to \cref{eq:mergepot}, we can rewrite \cref{eq:modirhoex} as 
\begin{equation}\label{eq:modirhoex2}
\rho^l_{\text{mid}} =  \sum_{i} \varrho^{0}_{i}
+ \sum_{k=1}^{l-1} \sum_{i}\sigma^{k}_{i}  + \rho^{l-1}_\text{int}, \quad \text{for} \quad l\geq 1,
\end{equation}
where $\varrho^{0}_{i}$ indicates the screening density induced by $[\Phi^0_i]$ at the boundaries of the $i$-th unit at Level 0, and $\sigma^{k}_{i}$ denotes the screening density induced by $\Psi^k_i$ at the boundaries of the $i$-th unit at Level $k\;(\geq1)$. Note that the two summations over $i$ in \cref{eq:modirhoex2} should be carried out only for the units in contact with the middle line at Level $l$.

Since $\rho^l_\text{mid}$ gives rise to $\Psi^l$, which induces $\sigma^l$, one can calculate $\sigma^{l}$ using 
\begin{equation}\label{eq:d2d}
\mathcal{K}^l_\text{D2D} \rho^l_\text{mid} = \sigma^l, \quad \text{for} \quad l\geq 1,
\end{equation}
where $\mathcal{K}^l_\text{D2D}$ is a density-to-density (D2D) kernel at Level $l$. 
The elements of $\mathcal{K}^l_\text{D2D}$ are readily computed by setting up a single cell with a unity density along the middle line and evaluating $\Psi_l$ at the cells in contact with the boundaries using the method of \citet{Muller19}. Note that the screening density $\sigma^l$ contributes to $\rho^k_\text{mid}$ for $k>l$. Thus, we first use \cref{eq:rholine} to compute $\rho_\text{mid}^1$ at Level 1, then successively solve \cref{eq:d2d,eq:modirhoex2}  from $l=1$ to the highest level. The screening density at the boundaries of the entire domain is given by 
\begin{equation}\label{eq:modirhoex3}
\varrho^{L} =  \sum_{i} \varrho^{0}_{i}
+ \sum_{k=1}^{L} \sum_{i} \sigma^{k}_{i},
\end{equation}
where $L=n_r+n_\theta-4$ denotes the highest level number, and the summations over $i$ include only the units touching the boundaries. \autoref{sec:DGF} describes the calculation of the open-boundary potential generated by $\varrho^L$.

\subsection{Case of $N_r=2^{n_r}$ and $N_\theta=2^{n_\theta}$ }\label{sec:power}

So far, we have explained the divide-and-conquer method for calculating the zero-boundary potential of a given density distribution when the simulation domain is discretized into a $(2^{n_r}-1)\times (2^{n_\theta}-1)$ mesh in the radial and polar directions. However, most simulations employ a mesh with power-of-two resolutions: $N_r=2^{n_r}$ and $N_\theta=2^{n_\theta}$.  In the following, we describe how to adapt the algorithm to handle such grid sizes while maintaining its efficiency and accuracy.

\begin{figure} 
	\centering
    \includegraphics[width=9cm]{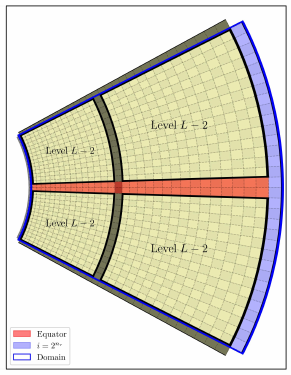}
    \caption{Structures of units and boundaries when the domain is discretized using a mesh with $N_r=2^{n_r}$ and $N_\theta=2^{n_\theta}$. The four units at Level $L-2$ have $15\times 15$ cells when $n_r=n_\theta=5$. The two polar grids in contact with the equator correspond to the lower and upper boundaries of the upper and lower units, respectively, while the outermost radial grid with $i=2^{n_r}$ coincides with the outer boundaries of the outer units in the radial direction.
    \label{fig:Ngrid}}
\end{figure}

Maintaining equatorial symmetry is beneficial as it allows the same kernel to be applied to both the northern and southern hemispheres, effectively halving the total kernel size. We configure the units in our divide-and-conquer scheme to exclude the outermost radial grid and the two polar grids touching the equator, while preserving equatorial symmetry in the polar direction. \cref{fig:Ngrid} illustrates the case of $n_r=n_\theta=5$, showing four units with $(2^{n_r-1}-1)\times (2^{n_\theta-1}-1)$ cells and associated boundaries at the third-highest level. Note that the units in the northern and southern hemispheres do not share common boundaries and that the outermost radial grid with $i=2^{n_r}$ coincides with the boundaries of the two outer units. These units are first merged in the polar direction and then in the radial direction to construct the entire domain at the highest level. 
The question then becomes how to incorporate the excluded cells near the equator and at the outer radial boundary to obtain the potential for the entire domain.

We first explain how to account for the two polar grids near the equator. Our goal is to merge the units of size $(2^{n_r-1}-1)\times (2^{n_\theta-1}-1)$ at Level $L-2$ in the polar direction, along with the two polar grids near the equator, to construct a unit of size $(2^{n_r-1}-1)\times 2^{n_\theta}$ at Level $L-1$. Since there is no common boundary between the units, instead of \cref{eq:modirhoex}, we define the line densities at the two polar grids as
\begin{equation}\label{eq:Nrho}
\begin{split}
\rho^{L-1}_{\text{mid,north}} &= \varrho^{L-2}_\text{north} + \rho^{L-2}_{\text{int},\text{north}}, \\
\rho^{L-1}_{\text{mid,south}} &= \varrho^{L-2}_\text{south} + \rho^{L-2}_{\text{int},\text{south}},
\end{split}
\end{equation}
where the subscripts `north' and `south' refer to the polar grids just above and below the equator, respectively, and $\varrho^{L-2}$ indicates the screening density induced by $[\Phi^{L-2}]$. These line densities near the equator affect the screening density $\sigma^{L-1}$, the middle-line potential $\Psi_\text{mid}^{L-1}$, and the source-free potential $\mathcal{L}_\text{mid}^{L-1}$ in the domain of size $(2^{n_r-1}-1)\times 2^{n_\theta}$. \autoref{sec:twolines} describes the modifications to the related \cref{eq:d2d,eq:d2p,eq:p2p}.

For the radial direction, we follow a simpler approach. The two units at Level $L-1$ are merged in the radial direction to form the unit with size $(2^{n_r}-1)\times 2^{n_\theta}$ at the highest level, which does not contain the outermost grid at $i=2^{n_r}$ in the simulation domain. Due to the linearity of the Poisson equation, the total potential of the entire domain can be obtained by summing the potential of the largest unit and the potential of the outermost grid with density $\rho_{N_R}$. This can be efficiently achieved by adding $\rho_{N_R}$ to the screening density $\varrho^L$ in \cref{eq:modirhoex3} at the outer radial boundary of the largest unit.

\section{Potential due to the Screening Density}\label{sec:DGF}

The open-boundary potential $\Theta$ induced by the screening density $\varrho^L$ at the highest level can be obtained by convolving $\varrho^L$ with the DGF. The DGF, $\mathcal{G}_{i,i',j,j',k-k'}$, in spherical polar coordinates is the gravitational potential per unit mass due to a discrete point mass at $(i',j',k')$, satisfying 
\begin{equation}\label{eq:DGFori}
(\Delta^{2}_{r} + \Delta^{2}_{\theta} + \Delta^{2}_{\phi})
\mathcal{G}_{i,i',j,j',k-k'} = 4\pi G 
\frac{\delta_{ii'}\delta_{jj'}\delta_{kk'}}{V_{i',j',k'}},
\end{equation}
where the symbol $\delta_{ii'}$ is the Kronecker delta and $V_{i',j',k'}$ is the volume of the $(i',j',k')$-th cell. The Fourier transform of \cref{eq:DGFori} in the azimuthal direction results in 
\begin{equation}\label{eq:DGFFou}
(\Delta^{2}_{r} + \Delta^{2}_{\theta} + \lambda^m_{i,j})
\mathcal{G}^m_{i,i',j,j'} = 4\pi G 
\frac{\delta_{ii'}\delta_{jj'}}{V_{i',j'}},
\end{equation}
where $\mathcal{G}^m_{i,i',j,j'}$ is the Fourier transform of $\mathcal{G}_{i,i',j,j',k-k'}$. Here, we omit the index $k'$ in $V_{i',j',k'}$ because the grid spacing is uniform in the azimuthal direction. From \cref{eq:DPE,eq:DGFFou}, it follows that the Fourier-transformed open-boundary potential is given by
\begin{equation}\label{eq:DGFF}
\Phi^m_{i,j} =  \sum^{N_r}_{i'=1} \sum^{N_\theta}_{j'=1} 
\mathcal{G}^m_{i,i',j,j'} \rho^m_{i',j'} V_{i',j'}.
\end{equation} 
Note that $\mathcal{G}^m_{i,i',j,j'}$ is a five-dimensional tensor, making it computationally inefficient to use \cref{eq:DGFF} directly to compute $\Phi^m_{i,j}$ over the entire domain. Instead, we apply it only at the domain boundaries.

To evaluate $\mathcal{G}^m_{i,i',j,j'}$, we follow the method of \citet{Moon} and assume that the DGF asymptotically matches the continuous Green's function (CGF), $\mathcal{G}(\mathbf{x},\mathbf{x}')=-G/|\mathbf{x}-\mathbf{x}'|$, at large distances from the source. 
That is, we impose the boundary condition for solving \cref{eq:DGFFou} as 
\begin{equation}\label{eq:BC}
\mathcal{G}^m_{i,i',j,j'} \approx -G\sum^{N_\phi}_{k=1} \frac{ \cos \left( {2\pi m k}/{N_\phi} \right)}{\sqrt{r_i^2 + r_{i'}^2 - 2r_i r_{i'} \cos \theta_j \cos \theta_{j'} - 2r_i r_{i'} \sin \theta_j \sin \theta_{j'} \cos (2\pi k/N_\phi)}} \quad 
\text{(far from the source)}.
\end{equation}
To ensure that the boundaries are sufficiently distant from the source when computing the DGF, we add an extra mesh with $n_{\text{pad},r} = 16$ cells to the inner and outer radial boundaries, and $n_{\text{pad},\theta} = 4$ cells to the upper and lower polar boundaries. 
Similar to the approach used for obtaining various kernels described in \autoref{sec:method}, we impose a unit density confined to a single cell along the boundaries and solve \cref{eq:DGFFou} for the DGF exclusively at the boundaries of the original computational domain.

Once the DGF is determined, we use \cref{eq:DGFF} to compute the open-boundary potential $\Theta^m$ induced by the Fourier-transformed, screening density $\varrho^{m,L}$ at the highest level. The double summations in \cref{eq:DGFF} can be reduced to a single summation over the boundary cells. The open-boundary potentials at the top, bottom, left, and right boundaries are then given by
\begin{align}
  \begin{split}\label{eq:DGFfulls}
\Theta^m_i (\text{top}) &= 
\sum^{N_r}_{i'=1} \tilde{\mathcal{G}}^m_{i,i',N_\theta+1,N_\theta+1} \varrho^{m,L}_{i',N_\theta+1} +
\sum^{N_r}_{i'=1} \tilde{\mathcal{G}}^m_{i,i',N_\theta+1,0} \varrho^{m,L}_{i',0} \\ &\phantom{skip}+
\sum^{N_\theta}_{j'=1} \tilde{\mathcal{G}}^m_{i,0,N_\theta+1,j'} \varrho^{m,L}_{0,j'} +
\sum^{N_\theta}_{j'=1} \tilde{\mathcal{G}}^m_{i,N_r+1,N_\theta+1,j'} \varrho^{m,L}_{N_r+1,j'}, 
\end{split}\\[1ex]
\begin{split}
\Theta^m_i  (\text{bottom}) &= 
\sum^{N_r}_{i'=1} \tilde{\mathcal{G}}^m_{i,i',0,N_\theta+1} \varrho^{m,L}_{i',N_\theta+1} +
\sum^{N_r}_{i'=1} \tilde{\mathcal{G}}^m_{i,i',0,0} \varrho^{m,L}_{i',0} \\&\phantom{skip}+
\sum^{N_\theta}_{j'=1} \tilde{\mathcal{G}}^m_{i,0,0,j'} \varrho^{m,L}_{0,j'} +
\sum^{N_\theta}_{j'=1} \tilde{\mathcal{G}}^m_{i,N_r+1,0,j'} \varrho^{m,L}_{N_r+1,j'}, 
\end{split}\\[1ex]
\begin{split}
\Theta^m_j  (\text{left}) &= 
\sum^{N_r}_{i'=1} \tilde{\mathcal{G}}^m_{0,i',j,N_\theta+1} \varrho^{m,L}_{i',N_\theta+1} +
\sum^{N_r}_{i'=1} \tilde{\mathcal{G}}^m_{0,i',j,0} \varrho^{m,L}_{i',0} \\&\phantom{skip} +
\sum^{N_\theta}_{j'=1} \tilde{\mathcal{G}}^m_{0,0,j,j'} \varrho^{m,L}_{0,j'} +
\sum^{N_\theta}_{j'=1} \tilde{\mathcal{G}}^m_{0,N_r+1,j,j'} \varrho^{m,L}_{N_r+1,j'}, 
\end{split}\\[1ex]
\begin{split}\label{eq:DGFfulle}
\Theta^m_j  (\text{right})&= 
\sum^{N_r}_{i'=1} \tilde{\mathcal{G}}^m_{N_r+1,i',j,N_\theta+1} \varrho^{m,L}_{i',N_\theta+1} +
\sum^{N_r}_{i'=1} \tilde{\mathcal{G}}^m_{N_r+1,i',j,0} \varrho^{m,L}_{i',0} \\&\phantom{skip}+
\sum^{N_\theta}_{j'=1} \tilde{\mathcal{G}}^m_{N_r+1,0,j,j'} \varrho^{m,L}_{0,j'} +
\sum^{N_\theta}_{j'=1} \tilde{\mathcal{G}}^m_{N_r+1,N_r+1,j,j'} \varrho^{m,L}_{N_r+1,j'},
  \end{split}   
\end{align}
where, for shorthand notation, we define $\tilde{\mathcal{G}}^m_{i,i',j,j'} \equiv \mathcal{G}^m_{i,i',j,j'} V_{i',j'}$. The four terms in the right-hand side of \Crefrange{eq:DGFfulls}{eq:DGFfulle} represent, from left to right, the contribution of the screening densities at the top, bottom, left and right boundaries. 

The four boundary potentials given in \Crefrange{eq:DGFfulls}{eq:DGFfulle} serve as the boundary values $\Theta_b^m$ for the interior potential $\Theta^m$ within the largest unit. Let $\mathcal{Z}_b^L$ denote the four boundary potentials. Noting that $\Theta^m$ is source-free within the unit, we follow the division process described in \autoref{sec:Zerodens}. Specifically, we solve the following equations
\begin{equation}\label{eq:DGF:p2p}
\mathcal{K}_\text{P2P}^l \mathcal{Z}_b^l = \mathcal{Z}_{\text{mid}}^{l}, \quad \text{for} \;\;  L\leq l \leq 1 ,\qquad \text{and}\qquad 
\mathcal{B}_\text{P2P} \mathcal{Z}_b^0 = \mathcal{Z}^{0}
\end{equation} 
to obtain the source-free potentials $\mathcal{Z}_{\text{mid}}^{l}$ at the middle line of each unit at Level $l$ and $\mathcal{Z}^{0}$ of each unit at the lowest level,
subject to the boundary condition $\mathcal{Z}_b^L$ imposed at the highest level.
Then, $\Theta^m$  is calculated as
\begin{equation}\label{eq:Phib_m}
\Theta^m = \sum_{i=1}^{2^L} \mathcal{Z}_i^{0}
+\sum_{l=1}^{L} \sum_{i=1}^{2^{L-l}}\mathcal{Z}_{\text{mid},i}^{l},
\end{equation} 
where the azimuthal mode $m$ is omitted on the right-hand side for simplicity. Note that $\Theta^m$ in \cref{eq:Phib_m} takes the same form as $\Psi^L$ in \cref{eq:Psi_sum} if $\Psi_\text{mid}^L$ is identified with $\mathcal{Z}_\text{mid}^L$. 

Combining \cref{eq:Pot_psi,eq:Phib_m}, we write the (Fourier-transformed) open-boundary potential as
\begin{equation}\label{eq:FinalPot_psi}
\Phi^m =\sum_{i=1}^{2^L} [\Phi^0_i] 
+ \sum_{l=1}^{L} \sum_i \mathcal{L}^0_i(l) 
+ \sum_{l=1}^{L} \sum^{l-1}_{k=1} \sum_i \mathcal{L}^k_{\text{mid},i}(l)  
+ \sum_i \mathcal{Z}^0_i 
+ \sum_{l=1}^{L} \sum_i \mathcal{Z}^l_{\text{mid},i}
+ \sum_{l=1}^{L} \sum_i \Psi^l_{\text{mid},i},
\end{equation}
where $\mathcal{L}^0(l)$ and $\mathcal{L}_\text{mid}^k(l)$ are the source-free boundary potentials at Level 0 and $k$, respectively, generated during the division phase in the computation of $\Psi^l$. Since $\mathcal{L}^0$ and $\mathcal{Z}^0$ share the same 
$\mathcal{B}_\text{P2P}$ kernels, and 
$\mathcal{L}_\text{mid}^l$ and $\mathcal{Z}_\text{mid}^l$ share the same $\mathcal{K}_\text{P2P}$ kernels as shown in \cref{eq:p2p,eq:p2p_0,eq:DGF:p2p}, their contributions in \cref{eq:FinalPot_psi} can be efficiently computed by employing the optimization technique described in \autoref{sec:optim}. As $\Phi^m$ is the solution to \cref{eq:DPE}, its inverse Fourier transform yields the desired open-boundary potential $\Phi$.

\section{Overall Procedure and Implementation}\label{sec:OverannImple}

In this section, we outline the overall procedure of our method and its computational implementation for computing the open-boundary potential $\Phi$ from the density distribution $\rho$.

\subsection{Algorithm Overview}\label{sec:Overview}

The key computational steps for solving the Poisson equation, along with their associated computational complexities, are summarized as follows.

\begin{enumerate}
    \item Perform a forward FFT of $\rho$ in the azimuthal direction to obtain $\rho_{i,j}^m$ (\autoref{eq:FFT}), which requires computational cost on the order of $\mathcal{O}(N_r N_\theta N_\phi \log N_\phi)$.

  \item Compute the zero-boundary potential $[\Phi^0]$ for all units at Level 0 (\autoref{eq:matrix1}), which has a computational cost of $\mathcal{O}(N_r N_\theta N_\phi)$.

 \item Compute the middle-line density $\rho^l_\text{mid}$ and screening density $\sigma^l$ at the boundaries of units at level $l$ by iteratively solving \cref{eq:modirhoex2,eq:d2d} from $l=1$ to the highest level $L$ using the D2D kernels. \autoref{sec:step3} shows that this requires a computational cost of $\mathcal{O}[N_r N_\theta N_\phi \log (N_\theta N_r)]$.
 
 \item Compute the screening density $\varrho^L$ at the boundaries of the entire domain using \cref{eq:modirhoex3}, which requires a computational cost of $\mathcal{O}[(N_r + N_\theta)N_\phi \log(N_rN_\theta)]$.

\item Compute the boundary potential $\Theta^m_b$ of the entire domain using the DGF and $\varrho^L$, as described in \Crefrange{eq:DGFfulls}{eq:DGFfulle}, which incurs a computational cost of $\mathcal{O}[(N_r + N_\theta)^2N_\phi]$ (see \autoref{sec:step7}).

\item Compute the zero-boundary potential $\Psi_\text{mid}^l$ due to $\rho_\text{mid}^l$ at the middle line of all units at Level $l\geq1$ using the D2P kernels (\autoref{eq:d2p}). This incurs a computational cost of $\mathcal{O}[N_r N_\theta N_\phi \log (N_\theta N_r)]$ (see \autoref{sec:step5}).

\item Compute $\mathcal{L}_\text{mid}^l$, $\mathcal{L}^0$, $\mathcal{Z}_\text{mid}^l$, $\mathcal{Z}^0$ using the P2P kernels in \cref{eq:p2p,eq:p2p_0,eq:DGF:p2p}. Then, calculate the Fourier-transformed open-boundary potential $\Phi_{i,j}^m$ from \cref{eq:FinalPot_psi}. This involves a computational cost of $\mathcal{O}[N_r N_\theta N_\phi \log (N_\theta N_r)]$ (see \autoref{sec:step6}).

\item Perform an inverse FFT of $\Phi_{i,j}^m $ along the azimuthal direction to obtain the open-boundary potential $\Phi$, which incurs a computational cost of $\mathcal{O}(N_r N_\theta N_\phi \log N_\phi)$.    
\end{enumerate}

Overall, the dominant computational cost arises from Steps 3, 6, and 7, which involve D2D, D2P, and P2P kernel matrices. As a result, the total computational cost of our Poisson solver scales as $\mathcal{O}(N^{3} \log N)$ for large $N=N_r \sim N_\theta \sim N_\phi$.

\subsection{Implementation}

We implement our Poisson solver in \texttt{FARGO3D} \citep{bm2016} using spherical polar coordinates. To enhance computational efficiency, we precalculate various D2P, P2P, and D2D kernels during initialization. To fully exploit the symmetric properties of the kernels, it is preferable to use logarithmic spacing in the radial direction and enforce equatorial symmetry in the polar direction.

Several open-source libraries are utilized for the implementation of the algorithm. The \texttt{fftw3} library \citep{FFTW} is used for performing the FFT. Since the \texttt{FARGO3D} code does not partition the domain in the azimuthal direction, MPI communication is unnecessary in that direction. Instead, we use the functions \texttt{fftw\_plan\_dft\_r2c\_1d()} and \texttt{fftw\_plan\_dft\_c2r\_1d()}, which are suitable for single-core FFT computations. The $\texttt{gsl}$ library\footnote{\url{https://www.gnu.org/software/gsl/}} is utilized to compute the inverse matrix and eigenvectors necessary for constructing the kernel matrices. For efficient multiplication of kernel matrices with densities or potentials, the $\texttt{BLAS}$ routine\footnote{\url{https://www.netlib.org/blas/}} is employed to achieve high-performance computations.

\begin{figure} 
	\centering
    \includegraphics[width=17cm]{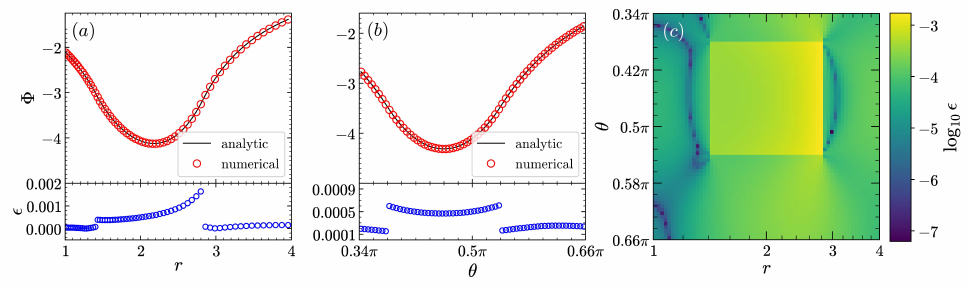}
    \caption{Test results for a uniform slender spherical wedge. The numerical potential (circles) is compared with the analytical potential (line) along ($a$) the $r$-direction at $\theta=0.5\pi$ and $\phi=0.125\pi$, and ($b$) the $\theta$-direction at $r=2$ and $\phi=0.125\pi$.  The lower panels in ($a$) and ($b$) display the corresponding cut profiles of the relative errors.  ($c$) Distribution of the relative errors on a logarithmic scale in the $\phi = 0.125\pi$ plane.
    \label{fig:acc1}}
\end{figure}

The volume of data and the frequency of inter-core communication are critical factors that impact the overall performance and efficiency of parallel computations. \autoref{sec:Parallel} describes how we partition the input and output variables for the kernel operations given in \cref{eq:d2p,eq:p2p,eq:d2d} to optimize communication efficiency. In short, there is a critical grid level in our hierarchical grid scheme where the grid size matches the size of a meshblock assigned to a single core. At lower levels, we partition the variables in the radial and polar directions, and in the azimuthal direction at higher levels. The inter-core communication occurs only at two points, involving small data sizes: during the bottom-up construction of the screening density and the top-down construction of boundary potential.
The theoretical communication cost per core of our approach is $8N_{\phi} (N_{r} N_\theta /N_\text{core})^{1/2}$, where $N_\text{core}$ is the number of cores used, This is significantly smaller than the communication required for MHD calculations.

\section{Test Results}\label{sec:tests}

We validate our solver through a series of test problems to assess its accuracy, convergence, and parallel performance. Additionally, we conduct time-dependent simulations of a gravitationally unstable isothermal ring to ensure that the gravity module integrates correctly with the MHD solver in \texttt{FARGO3D}, reproducing results consistent with those obtained from another validated code. For all tests presented below, we set the gravitational constant to $G=1$.

\subsection{Accuracy Test}\label{sec:accuracy}

To test the accuracy of our spherical Poisson solver, we design a solid figure whose shape matches the cell shape of the adopted grid and remains unchanged with varying resolution. For this purpose, we consider a uniform slender spherical wedge with a density of unity, occupying the region $[\sqrt{2}, 2\sqrt{2}]\times[0.38\pi, 0.54\pi]\times [0,0.25\pi]$ within a domain of size $[1,4]\times[0.34\pi, 0.66\pi]\times [0,2\pi]$. While no explicit algebraic expression exists for the potential of this figure, \citet{Ziegler24} provided a closed-form formulation in their appendix, expressed in terms of line and surface integrals \citep[see also,][]{Hure}. We evaluate this integral-form solution, using the Romberg integration method with an accuracy of $10^{-8}$ to obtain the reference potential $\Phi_r$.

\begin{figure} 
	\centering
    \includegraphics[width=10cm]{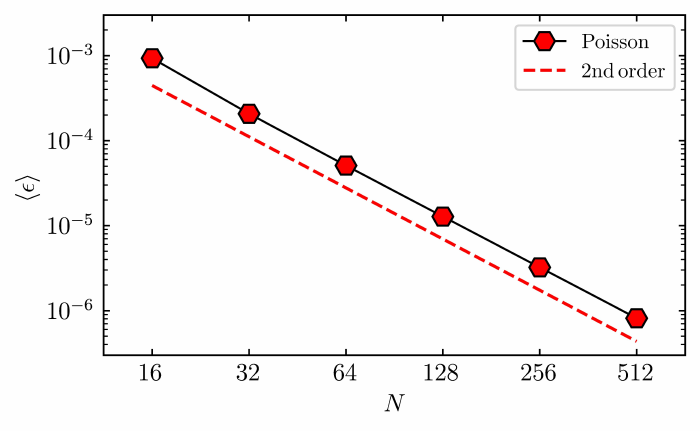}
    \caption{Convergence test results for the Poisson solver on the uniform slender spherical wedge with a density of unity, as described in \autoref{sec:accuracy}. The computational domain is set to $N \times N \times 4N$ for each $N$. The dashed line represents the ideal convergence slope of $-2$.
    \label{fig:conv}}
\end{figure}

We discretize the computation domain into $N\times N\times 4N$ cells with $N=64$, and calculate the open-boundary potential $\Phi$ of the figure using our algorithm. To assess accuracy, we define the relative error between the numerical and reference solutions as
\begin{equation}\label{eq:error}
\epsilon \equiv  \Bigg| \frac{\Phi - \Phi_{r}}{\Phi_{r}} \Bigg|,
\end{equation}
evaluated over the entire domain. \cref{fig:acc1} plots cut profiles of $\Phi$ and $\epsilon$ along the radial direction at $\theta=0.5\pi$ and $\phi=0.125\pi$, and along the polar direction at $r=2$ and $\phi=0.125\pi$, as well as the in-plane distribution of $\epsilon$ at $\phi=0.125\pi$. The relative errors remain very small, at approximately $0.1\%$, demonstrating excellent agreement with the analytical solution.

\subsection{Convergence Test}\label{sec:convergence}

Since the Laplace operators in \cref{eq:DPE} are second-order accurate, doubling the resolution is expected to improve the accuracy of the calculated potential by approximately a factor of 4. We evaluate the performance of our self-gravity solver by varying the grid resolution $N$ from $16$ to $512$ for the same solid figure described in \autoref{sec:accuracy}. \cref{fig:conv} plots the averaged relative error $\langle\epsilon\rangle$ as a function of $N$ over the range $[2^4, 2^9]$. Linear regression of the numerical results gives a slope of $-2.023$, confirming that our Poisson solver implementation achieves second-order accuracy.

\subsection{Performance Test}

To evaluate the performance of our Poisson solver, we use the \texttt{Stellar} cluster at Princeton University, utilizing compute nodes equipped with Intel Cascade Lake CPUs. We conduct a weak scaling test to analyze how computation speed scales with the number of cores, with each core handling grids of equal size. To enhance efficiency and parallelization, the \texttt{FARGO3D} code avoids partitioning along the azimuthal direction and adopts a square-shaped meshblock in the radial and polar directions. In our test, we allocate a meshblock of size $16 \times 16 \times 1024$ to each core.  We then measure the wallclock time per cycle by averaging the times recorded over 100 cycles, using core counts $N_\text{core}$ of 16, 64, 256, 1024, and 4096. For $N_\text{core} = 4096$, the total grid size is $(N_r, N_\theta, N_\phi)=(1024, 1024, 1024)$.
Additionally, we compare the computation speed of the gravity solver to that of the MHD computation for each core count.

\begin{figure} 
	\centering
    \includegraphics[width=10cm]{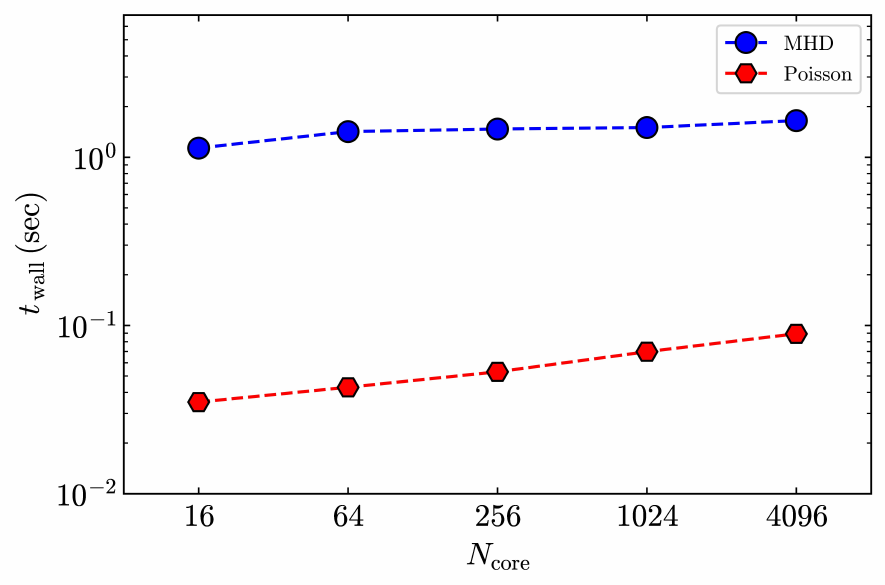}
    \caption{Results of the weak scaling test. The wallclock time per cycle, $t_\text{wall}$, of the Poisson solver is compared with that of the MHD solver in \texttt{FARGO3D} as a function of the number of processors, $N_\text{core}$.
    \label{fig:weak}}
\end{figure}

\Cref{fig:weak} plots the wallclock time per cycle, $t_\text{wall}$, for the gravity solver, compared with that of the MHD solver, as a function of $N_\text{core}$. 
\Cref{tab:weakscaling} presents the 
corresponding quantitative values of $t_\text{wall}$ and the update time per cell, $t_\text{update} = t_\text{wall} / (16\times 16\times 1024)$, for both the Poisson solver and the MHD solver. Ideally, $t_\text{wall}$ should remain independent of $N_\text{core}$; however, our experiments show that $t_\text{wall}$ increases by nearly a factor of 2.5 when $N_\text{core}$ is increased by a factor of 256. This result is remarkable, considering that the gravity solver requires global communication, which usually scales poorly at high core counts. Despite this, the observed increase in wallclock time is relatively modest, suggesting that the solver maintains a reasonable level of scalability. The deviation from ideal scaling is likely due to communication overhead and load imbalance, which become more pronounced as the number of cores increases. Notably, the time taken for calculating self-gravity is only a few percent of that for the MHD solver in \texttt{FARGO3D}, indicating that the inclusion of self-gravity does not significantly impact the overall computational cost. This efficiency makes the solver well-suited for large-scale simulations where self-gravity is essential, such as studies of protoplanetary disks and galactic dynamics.

\begin{deluxetable}{@{\extracolsep{4pt}}ccccc}
\tablecaption{Results of the weak scaling test, listing the wallclock time per cycle and the update time per cell for both the Poisson solver and the MHD solver in \texttt{FARGO3D} for different numbers of cores.
\label{tab:weakscaling}}
\tablehead{
\colhead{} & \multicolumn{2}{c}{Wall clock time (s)} 
& \multicolumn{2}{c}{Update time per cell ($\mu$s)} \\
\cline{2-3} \cline{4-5}
\colhead{Number of Cores} &
\colhead{MHD} & \colhead{Poisson} &
\colhead{MHD} & \colhead{Poisson} 
}
\startdata
16 & 1.134 & 0.03503 & 4.325 & 0.1336\\
64 & 1.420 & 0.04292 & 5.418 & 0.1637\\
256 & 1.472 & 0.05300 & 5.616 & 0.2022\\
1024 & 1.502 & 0.06979 & 5.728 & 0.2662\\
4096 & 1.652 & 0.08929 & 6.304 & 0.3406\\
\enddata
\end{deluxetable}

\subsection{Ring Fragmentation}

As a final test, we perform a time-dependent simulation of the fragmentation of a self-gravitating isothermal ring using \texttt{FARGO3D}. We initially consider a rigidly rotating ring at angular velocity $\Omega_0$ and sound speed $c_s$, immersed in a hot, tenuous medium. The ring is initially in hydrosteady equilibrium, subject to both self-gravity and external gravity. Assuming that the external gravitational field, $\mathbf{g}_\text{ext}=-\Omega_e^2\sin\theta \mathbf{r}$, alone causes the ring to rotate at angular velocity $\Omega_e$, the equilibrium condition requires 
\begin{equation}\label{eq:HSE}
\boldsymbol{\nabla}\left(c_s^2 \ln \rho + \Phi\right) = \Omega_s^2 \sin\theta \mathbf{r},
\end{equation}
where $\Omega_s=(\Omega_0^2-\Omega_e^2)^{1/2}$ is the angular velocity due to self-gravity alone \citep{Ring16}. It can be shown that the equilibrium states are characterized by two dimensionless parameters: $\alpha \equiv c_s^2/(GR_A^2\rho_c)$ and $\hat{\Omega}_s\equiv\Omega_s/(G\rho_c)^{1/2}$, where $\rho_c$ and $R_A$ represent the maximum density and the maximum radial distance of the ring, respectively.

\begin{figure}
 \plotone{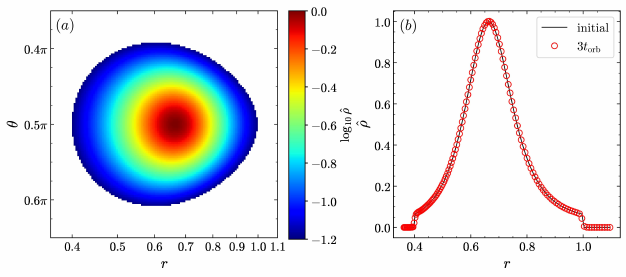}
  \caption{ $(a)$ Initial meridional distributions of the normalized ring density $\hat\rho =\rho/\rho_c$
  in the $\phi=0$ plane. 
  $(b)$ Comparison of the radial distribution of  $\hat\rho$ at $\phi=0$ and $\theta=\pi/2$ between $t/t_\text{orb}=0$ and $3$.}
  \label{fig:dyn1}
\end{figure}

For our test simulation, we configure a logarithmic spherical polar grid with a domain size of $r \in [0.36,1.1]$, $\theta \in [0.35\pi,0.65\pi]$, and $\phi \in [0,2\pi]$. The grid resolution is set to $(N_{r}, N_{\theta}, N_{\phi}) = (128, 128, 512)$. We use the self-consistent field method of \citet{Hachisu} to construct a ring with $\alpha = 3.0 \times 10^{-2}$ and $\hat{\Omega}_s = 0.47$. We then apply the external gravity to increase its rotational velocity to $\hat{\Omega}_0 \equiv \Omega_0/(G\rho_c)^{1/2}=0.5$. Without applying any perturbations, we first monitor how well the initial equilibrium configuration is maintained using the \texttt{FARGO3D} code. \cref{fig:dyn1} plots the meridional distribution of $\hat{\rho}\equiv\rho/\rho_c$ at the $\phi=0$ plane and compares the radial distribution of the ring density at $\theta=\pi/2$ and $\phi=0$ between $t/t_\text{orb}=0$ and $3$, where $t_\text{orb}=2\pi/\hat{\Omega}_0$. Clearly, the equilibrium configuration is well-maintained over time, with only slight diffusion occurring near the ring boundaries. This suggests that our Poisson solver accurately computes the self-gravity of the ring.

\begin{figure}
 \plotone{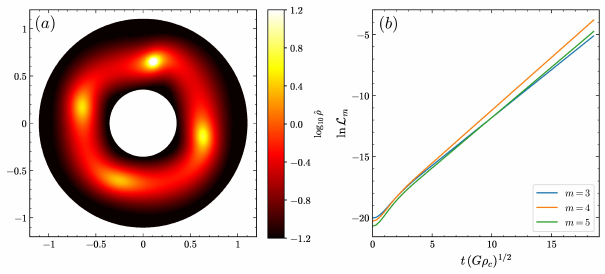} 
  \caption{$(a)$ Equatorial distribution of the ring density $\hat{\rho}$ at $t/t_\text{orb} = 19.2$.
  $(b)$ Temporal growth of the Fourier transform $\mathcal{L}_m$ of the integrated ring density for the modes with $m=3,4$, and $5$. The $m=4$ mode grows the fastest.}
  \label{fig:dyn2}
\end{figure}

Next, we introduce small-amplitude random density perturbations to the equilibrium configuration and allow the ring to evolve. We find that the perturbations grow exponentially with time, eventually resulting in the formation of four clumps, as shown in \cref{fig:dyn2}($a$). To quantify the growth rates of various modes in the simulation, we define the radially and vertically integrated density as $\mathcal{L}(\phi) \equiv \int \rho r^2 \sin \theta \,dr d\theta$ and compute the Fourier amplitude of mode $m$ as $\mathcal{L}_m = e^{i\vartheta_m} \int \mathcal{L}(\phi) e^{-im\phi} d\phi $, with $\vartheta_m$ being the phase angle. \cref{fig:dyn2}($b$) plots the growth of $\mathcal{L}_m$ for $m=3,4$, and 5. The initial random perturbations go through an early relaxation phase with $t(G\rho_c)^{1/2} \lesssim 3$, followed by an exponential growth phase of the self-gravitating instability. The growth rates of the modes are measured to be $0.786$, $0.866$, and $0.826$ for $m=3,4$, and 5, respectively.

Since there is no analytic solution for the self-gravitating instability of the ring, we replicate the same experiment using the cylindrical Poisson solver of \citet{Moon}, which is implemented in the \texttt{Athena++} code. It turns out that the results from the cylindrical solver are almost identical to those obtained with our solver, with the growth rates for the $m=3$, $4$, and $5$ modes differing only by $0.4\%$, $0.05\%$, and $0.6\%$, respectively, from the values reported above. This is remarkable considering the different grid geometries used in the two codes. 

The test results presented in this section collectively show that our three-dimensional spherical Poisson solver is dependable, second-order accurate, and highly efficient.

\section{Summary and Discussion}\label{sec:summary}

To study the dynamics of self-gravitating, rotating, and flared protoplanetary disks, it is crucial to implement a Poisson solver in spherical polar coordinates with a confined range of polar angles. In this paper, we present an accurate algorithm for a Poisson solver that is optimized for efficiency in logarithmic spherical coordinates.
Our algorithm employs a divide-and-conquer strategy, partitioning the computational domain into hierarchical units with varying cell sizes, and utilizes James's method to calculate the potential for each unit.

We begin by reducing the Poisson equation to obtain \cref{eq:DPE} through an FFT in the azimuthal direction. We then transform \cref{eq:DPE} into a series of kernel operations, as shown in \cref{eq:d2p,eq:p2p,eq:d2d}, employing D2P, P2P, and D2D kernels of varying sizes for all Fourier modes $m$. These kernel matrices are precomputed using the method described by \citet{Muller19} (see \autoref{sec:proj} and \autoref{sec:p2p}). Although the calculation of these kernel matrices demands significant computational effort, it only needs to be performed once at the beginning of the simulation. By carefully optimizing the kernel operations, we reduce the computational cost of determining the potential for a given density distribution to $\mathcal{O}(N^{3} \log N)$.

We implement our algorithm in $\texttt{FARGO3D}$ and conduct several test simulations to assess its accuracy, performance, and parallel efficiency. The accuracy is found to follow second-order convergence as the grid resolution increases, with the absolute accuracy also proving to be quite high. The algorithm is significantly faster than the MHD solver for various grid sizes, and parallel efficiency demonstrates strong results up to 1024 cores in weak scaling tests. A test simulation on ring fragmentation shows that our algorithm accurately captures the evolution of a time-dependent system.

\citet{Ziegler24} proposed a method for solving the internal potential using James's algorithm in combination with a multigrid approach. As mentioned in \autoref{sec:intro}, however, it appears that their self-gravity solver takes longer to execute than the MHD calculations. By avoiding partitioning in the azimuthal direction, the communication cost in our algorithm is optimized, enabling efficient potential calculation without encountering parallel inefficiencies as the number of cores increases.  Although an exact comparison is challenging due to differences in computing environments and the varying convergence rates of the multigrid method, our solver exhibits efficient computational speed and favorable scaling results.

\section*{acknowledgments}
We are grateful to the referee for constructive comments. We also acknowledge Dr.\ S.\ Moon for kindly providing his implementation of the cylindrical Poisson solver and for his assistance with the weak scaling tests. This work was supported by the National Research Foundation of Korea (NRF) grant funded by the Korea government (MSIT) (RS-2025-00517264), and by Korea Institute of Science and Technology Information (KISTI) under the institutional R\&D project (K25L2M2C3). Computational resources for this project were provided by the Supercomputing Center/KISTI with supercomputing resources including technical support (KSC-2024-CRE-0207); the Center for Advanced Computation at the Korea Institute for Advanced Study; Princeton Research Computing, a consortium including PICSciE and OIT at Princeton University.

\software{\texttt{FARGO3D}(\citealt{FARGO2}, \url{https://github.com/FARGO3D/fargo3d}), \texttt{fftw3}(\citealt{FFTW}, \url{https://www.fftw.org}), \texttt{gsl}(\url{https://www.gnu.org/software/gsl/}), \texttt{BLAS}(\url{https://www.netlib.org/blas/}) }

\appendix

\section{Projection Method}\label{sec:proj}

\citet{Muller19} introduced a Poisson solver utilizing matrix multiplication, which can be effectively applied to compute kernel matrices within our divide-and-conquer strategy. The core idea of their method is to determine a transformation matrix that diagonalizes the Laplace operators in both the polar and azimuthal directions. First, we rewrite \cref{eq:DPE} as 
\begin{equation}\label{eq:ProjDel}
\Delta^{2}_{r}\Phi^{m}_{i,j} + 
\mathcal{R}_i (\hat{\Delta}^2_\theta\Phi^{m}_{i,j} + 
\mathcal{S}_j \lambda^{m}_{\phi}\Phi^{m}_{i,j}) = 4\pi G \rho^{m}_{i,j},
\end{equation}
where 
\begin{equation}
\hat{\Delta}^2_\theta\Phi_{i,j}^m \equiv \frac{1}{\cos\theta_{j-1/2}-\cos\theta_{j+1/2}} \bigg( \frac{\Phi^m_{i,j+1}-\Phi^m_{i,j}}{\theta_{j+1}-\theta_{j}}\sin\theta_{j+1/2}
- \frac{\Phi^m_{i,j}-\Phi^m_{i,j-1}}{\theta_{j}-\theta_{j-1}}\sin\theta_{j-1/2} \bigg),
\end{equation}
with
\begin{equation}
\lambda^m_{\phi} \equiv-\left[ \sin\left(\frac{m\pi}{N_{\phi}} \right)\Big/ \frac{\pi}{N_{\phi}} \right]^{2}.
\end{equation}
Next, we define the transformation matrix $H^m_{l,j}$ that satisfies 
\begin{equation}\label{eq:Proj2}
\hat{\Delta}^2_\theta H^m_{l,j} + \mathcal{S}_j \lambda^{m}_{\phi} H^m_{l,j} = \lambda^m_{\Omega,l} H^m_{l,j},
\end{equation}
where $\lambda^m_{\Omega,l}$ is the eigenvalue of the operator $\hat{\Delta}^2_\theta + \mathcal{S}_j \lambda^{m}_{\phi}$. Projecting  \cref{eq:ProjDel} onto $H^m_{l,j}$ gives  
\begin{equation}\label{eq:MullerTri}
\Delta^{2}_{r} \hat{\Phi}^{m}_{i,l} + 
\mathcal{R}_i \lambda^m_{\Omega,l} \hat{\Phi}^{m}_{i,l} = 4\pi G \hat{\rho}^{m}_{i,l},
\end{equation}
where $\hat{\Phi}^{m}_{i,l}$ and $\hat{\rho}^m_{i,j}$ are projected matrices defined through 
\begin{equation}\label{eq:MullerTrans1}
\Phi^m_{i,j} = \hat{\Phi}^m_{i,l}H^m_{l,j},\quad \text{and} \quad 
\rho^m_{i,j} = \hat{\rho}^m_{i,l} H^m_{l,j}.
\end{equation}
Note that \cref{eq:MullerTri} forms a tridiagonal linear system in $i$ for given $m$ and $l$, which can be solved using Thomas algorithm.
Once \cref{eq:MullerTri} is solved with appropriate boundary conditions,  $\Phi^m_{i,j}$ can be computed using \cref{eq:MullerTrans1}. 

Let us briefly discuss the computational complexity of \citet{Muller19}'s method. 
For simplicity, consider the case with $N_r =N_\theta = N_\phi = N$. Solving the tridiagonal system using Thomas algorithm requires a computational cost of $\mathcal{O}(N)$. Consequently, solving \cref{eq:MullerTri} for all combinations of $m$ and $l$ results in a total cost of $\mathcal{O}(N^3)$. For a given $m$, the matrix multiplications in \cref{eq:MullerTrans1} scale as $\mathcal{O}(N^3)$, leading to an overall cost of $\mathcal{O}(N^4)$ for all $m$. Therefore, the projection method is computationally expensive as a Poisson solver for time-varying $\rho$, especially for large $N$.

We apply this projection method solely to derive kernels that relate the potential and density when only a single cell has nonzero density or potential. In this case, the matrix multiplications in \cref{eq:MullerTrans1} change to matrix-vector multiplications. The associated computational cost of finding a kernel of size $n\times n$ is $\mathcal{O}(n^3)$. Since lower-level kernels are small in size and can exploit radial similarities, the time to compute the entire kernels is dominated by the largest kernel. The size of the largest kernel is approximately $N \times N$, and the computation required to obtain it is $\mathcal{O}(N^3)$. Multiplying by $N_\phi$ to account for each azimuthal mode gives $\mathcal{O}(N^4)$. While this computation remains large, it is acceptable because the kernel computations are performed only once during the initialization phase of time-dependent simulations.

\section{Potential-to-Potential Kernel}\label{sec:p2p}

We introduce the P2P kernel $\mathcal{K}_\text{P2P}^l$ in \cref{eq:p2p} to compute the middle-line potential $\mathcal{L}_\text{mid}^l$ from the boundary potential $\mathcal{L}_b^l$ on a unit at Level $l$. All cells within the unit have zero density. 

To evaluate $\mathcal{K}_\text{P2P}^l$, we first consider a situation where only a single boundary cell has a potential of unity, while all other boundary cells have zero potential.  Let $\mathcal{L}_p$ represent the potential distribution at the boundaries, and let $\mathcal{L}$ denote the unknown internal potential within the unit induced by $\mathcal{L}_p$. Since $\boldsymbol{\nabla}^2\mathcal{L}=0$, the virtual density $\rho_p$ within the unit that generates $\mathcal{L}_p$ can be determined using \cref{eq:DPE}. 
This implies that $\mathcal{L} - \mathcal{L}_p$ is the zero-boundary potential corresponding to the internal density $-\rho_p$. By using the projection method of \citet{Muller19}, outlined in \autoref{sec:proj}, we calculate the zero-boundary potential along the middle line due to the internal density $-\rho_p$. Inserting the resulting potential into \cref{eq:p2p} for $\mathcal{L}_p$ provides the elements in a column of $\mathcal{K}_\text{P2P}^l$. By repeating the same procedure for different boundary cells, we populate all elements of $\mathcal{K}_\text{P2P}^l$. 

\section{Handling of Two-Line Densities Near the Equator}\label{sec:twolines}

In this section, we discuss the modifications to the equations used to calculate the screening density $\sigma^{L-1}$, the middle-line potential $\Psi_\text{mid}^{L-1}$, and the source-free potential $\mathcal{L}_\text{mid}^{L-1}$ caused by the two-line densities given in \cref{eq:Nrho}. These line densities near the equator induce the line potential $\Psi^{L-1}$, which in turn generates $\sigma^{L-1}$ at the boundaries of the domain of size $(2^{n_r-1}-1)\times 2^{n_\theta}$ at Level $L-1$. Using the D2D kernel matrix, this relationship can be expressed as
\begin{equation}\label{eq:tildaD2D}
\mathcal{K}^{L-1}_\text{D2D} \rho^{L-1}_{\text{mid,north}} +
(\mathcal{K}^{L-1}_\text{D2D}
\rho^{L-1}_{\text{mid,south}})^* = \sigma^{L-1},
\end{equation}
which replaces \cref{eq:d2d}. Here, the asterisk indicates reflection about the equator: the $\mathcal{K}^{L-1}_\text{D2D}$ kernel has a size of $(2^{n_r}+2^{n_\theta+1}-2)\times (2^{n_r-1}-1)$ and is defined for the north line.

The two-line densities also affect the middle-line potential, $\Psi_\text{mid}^{L-1}$, and the source-free potential, $\mathcal{L}_\text{mid}^{L-1}$. Consequently, \cref{eq:d2p,eq:p2p} should be modified at Level $L-1$ to 
\begin{equation}\label{eq:tildaD2P}
\mathcal{K}^{L-1}_\text{D2P} \rho^{L-1}_\text{mid,north} + 
(\mathcal{K}^{L-1}_\text{D2P}
\rho^{L-1}_\text{mid,south})^* = \Psi^{L-1}_{\text{mid},\text{north}} + \Psi^{L-1}_{\text{mid},\text{south}}, 
\end{equation}
and
\begin{equation}\label{eq:tildaP2P}
\mathcal{K}^{L-1}_\text{P2P}\mathcal{L}^{L-1}_b = \mathcal{L}^{L-1}_{\text{mid},\text{north}}, \quad
\mathcal{K}^{L-1}_\text{P2P}(\mathcal{L}^{L-1}_b)^* = \mathcal{L}^{L-1}_{\text{mid},\text{south}}, 
\end{equation}
where $\Psi^{L-1}_{\text{mid},\text{north}}$ and $\Psi^{L-1}_{\text{mid},\text{south}}$ denote the potentials generated by $\rho^{L-1}_\text{mid,north}$ and $\rho^{L-1}_\text{mid,south}$ in the lines just above and below the equator, respectively, while $\mathcal{L}^{L-1}_{\text{mid},\text{north}}$ and $\mathcal{L}^{L-1}_{\text{mid},\text{south}}$ represent the source-free potentials in the same regions. Again, the asterisks denote reflection relative to the equator. The sizes of $\mathcal{K}^{L-1}_\text{D2P}$ and $\mathcal{K}^{L-1}_\text{P2P}$ are
$(2^{n_r}-2)\times (2^{n_r-1}-1)$ and
$(2^{n_r-1}-1)\times
(2^{n_r}+2^{n_\theta+1}-2)$, respectively.
For the upper unit at Level $L-2$, $\mathcal{L}_b^{L-2}=\Psi_{\text{mid},\text{north}}^{L-1}$ in the line just above the equator, and $\mathcal{L}_b^{L-2}=0$ at the other three boundaries. For the lower unit, similarly, $\mathcal{L}_b^{L-2}=\Psi_{\text{mid},\text{south}}^{L-1}$ in the line just below the equator, and $\mathcal{L}_b^{L-2}=0$ at the other three boundaries. The potentials calculated from \cref{eq:tildaP2P} also serve as the boundaries for the source-free potentials at Level $L-2$.

\section{Computational Complexities}\label{sec:CC}

Here, we describe the computational complexities of Steps 3, 5, 6 and 7 presented in \autoref{sec:Overview}, involved in the computation of the gravitational potential.

\subsection{Step 3: Involving the D2D Kernel}\label{sec:step3}

To calculate $\rho_\text{mid}^l$ at level $l$, we iteratively solve \cref{eq:modirhoex2,eq:d2d} from level $l=1$ to the highest level. Since \cref{eq:modirhoex2} involves simple vector additions, the associated computational cost is dominated by the matrix-vector multiplication in  \cref{eq:d2d}. For simplicity, we consider the case of $n_r = n_\theta$. According to the unit merging rule, the middle line is parallel to the polar direction at even levels and to the radial direction at odd levels. 

When $l$ is odd, the sizes of $\rho_\text{mid}^l$ and $\sigma^l$ are $2^{(l+1)/2+1}-1$ and $2^{(l+1)/2+3}+2^{(l+1)/2+2}-4$, respectively. The corresponding D2D kernel matrix has a size of 
$\left[2^{(l+1)/2+3}+2^{(l+1)/2+2}-4\right]\times \left[2^{(l+1)/2+1}-1\right]$. In contrast, when $l$ is even, the D2D kernel matrix has $(2^{l/2+4}-4)\times (2^{l/2+2}-1)$ elements. Since there are $2^{L-l}$ units at Level $l$, the number of operations required to solve \cref{eq:d2d} at Level $l$ is 
\begin{equation}\label{eq:step3}
\begin{split}
2^{L-l} \times (2^{\frac{l+1}{2}+3} + 2^{\frac{l+1}{2}+2}-4) \times (2^{\frac{l+1}{2}+1}-1) &\approx 3\times 2^{L+4}, \quad \text{for  odd $l$}, \\
2^{L-l} \times (2^{\frac{l}{2}+4}-4) \times (2^{\frac{l}{2}+2}-1) &\approx 2^{L+6}, \quad  \;\;\,\quad \text{for  even $l$},
\end{split}
\end{equation}
independent of $l\gg1$. Since \cref{eq:d2d} should be solved sequentially from $l=1$ to $l=L(=n_r+n_\theta-4)$ for all azimuthal mode $m$, the total computational cost of Step 3 is approximately $N_\phi (n_r + n_\theta-4)2^{n_r + n_\theta + 2} \sim \mathcal{O}[N_r N_\theta N_\phi \log (N_r N_\theta)]$.

\subsection{Step 5: involving the DGF}\label{sec:step7}

\Crefrange{eq:DGFfulls}{eq:DGFfulle} calculate the open-boundary potential due to the Fourier-transformed screeing density $\varrho^{m,L}$ at the highest level. These equations imply that, for a given $m$, we need to have sixteen 3D arrays of the DGF: four of size $N_\phi \times N_r\times N_r$, eight of size $N_\phi \times N_r\times N_\theta$, and four of size $N_\phi \times N_\theta\times N_\theta$, where $N_r=2^{n_r}$ and $N_\theta=2^{n_\theta}$. However, both the equatorial symmetry and the Hermitian symmetry of the Fourier transform for a real function (see, e.g., \autoref{eq:lambdam}) significantly reduce the memory requirements. Moreover, the radial dependence of the DGF can be efficiently accounted for using the scale factor $\mathcal{R}_i$.  As a result, only twelve 3D arrays of the DGF are required: eight of size $N_\phi/2 \times N_r\times N_\theta/2$, and four of size $N_\phi/2 \times N_\theta\times N_\theta/2$. In a more compact form, \Crefrange{eq:DGFfulls}{eq:DGFfulle} can be expressed as 
\begin{equation}\label{eq:DGFcom}
\tilde{\mathcal{G}}^m \varrho^{m,L} = \Theta_b^m,
\end{equation}
where $\tilde{\mathcal{G}}^m$ is a square matrix with $2(N_r + N_\theta -1)$ elements along each direction, and both $\varrho^{m,L}$ and $\Theta_b^m$ are vectors of size $2(N_r + N_\theta -1)$. The total computational cost for the matrix-vector multiplication in \cref{eq:DGFcom} across all $m$ is $2(N_r + N_\theta -1)^2N_\phi \sim \mathcal{O}[(N_r + N_\theta)^2N_\phi]$.

\subsection{Step 6: Involving the D2P Kernel}\label{sec:step5}

To calculate $\Psi_\text{mid}^l$ at Level $l$, we apply the D2P kernel matrix $\mathcal{K}_\text{D2P}^l$ to $\rho^l_\text{mid}$ using \cref{eq:d2p}. As noted in \autoref{sec:step3}, the size of $\rho_\text{mid}^l$ is $2^{(l+1)/2+1}-1$ for odd $l$, and $\mathcal{K}_\text{D2P}^l$ is a square matrix of size $(2^{(l+1)/2+1}-1)\times (2^{(l+1)/2+1}-1)$. For even $l$, $\mathcal{K}_\text{D2P}^l$ has $(2^{l/2+2}-1)\times(2^{l/2+2}-1)$ elements. Since there are $2^{L-l}$ units at Level $l$, the cost in evaluating $\Psi_\text{mid}^l$ is given by 
\begin{equation}
\begin{split}
2^{L-l} \times (2^{\frac{l+1}{2}+1}-1)^2 &\approx
2^{L-l} \times 2^{l+3} = 2^{L+3},\quad \text{for  odd $l$},  \\
2^{L-l} \times (2^{\frac{l}{2}+2}-1)^2 &\approx 
2^{L-l} \times 2^{l+4} = 2^{L+4}, \quad \text{for  even $l$},  
\end{split}
\end{equation}
independent of $l$, which is $\sim4$--$6$ times smaller than the computational cost for the D2D kernel in \cref{eq:step3}.  Since \cref{eq:d2p} must be solved for all azimuthal mode $m$ from $l=1$ to $L$, the total computational complexity of Step 6 is $\mathcal{O}[N_r N_\theta N_\phi \log (N_r N_\theta)]$, which matches the complexity of Step 3. 

\subsection{Step 7: Involving the P2P Kernel}\label{sec:step6}

To obtain $\Psi^l$ from \cref{eq:Psi_sum},  $\mathcal{L}_\text{mid}^l$ and $\mathcal{L}^0$ must be computed sequentially from Level $L$ to $1$ using \cref{eq:p2p,eq:p2p_0}, which involve the P2P kernels $\mathcal{K}_\text{P2P}^l$ and $\mathcal{B}_\text{P2P}$. Note the similarity between the P2P and D2D kernels: the former is used to compute the middle-line potential from the boundary potential, while the latter is used to compute the boundary density from the middle-line density. This similarity suggests that the computational cost of calculating $\mathcal{L}_\text{mid}^l$ is $\mathcal{O}[N_r N_\theta N_\phi \log (N_r N_\theta)]$, which is the same as the complexity of computing $\rho_\text{mid}^l$ in Step 3.

\section{Data Partitioning for Kernel Operations}\label{sec:Parallel}

Essentially, our method transforms the reduced Poisson \cref{eq:DPE} into a series of kernel operations in \cref{eq:d2p,eq:p2p,eq:d2d}, utilizing D2P, P2P, D2D kernels of varying sizes for all Fourier modes $m$. Since the eigenvalue $\lambda_{i,j}^m$ in \cref{eq:lambdam} depends on $m$, the elements of the kernel matrices vary for different values of $m$. The performance of our solver therefore depends on how efficiently these kernel operations are executed at different levels. To achieve optimal computation, it is crucial to properly partition the input and output variables in the kernel operations, such as $\rho_\text{mid}^l$ and $\sigma_l$ in \cref{eq:d2d}.

We use two distinct approaches to partition the input and output variables based on the grid level. At lower levels, the variables are partitioned in the radial and polar directions, following the standard approach used in the \texttt{FARGO3D} code, whereas at higher levels, they are partitioned only in the azimuthal direction. To illustrate this, let us consider the case where a meshblock of size $c\times c \times N_\phi$ is assigned to a single core, with $c$ being a power of two. Let $l_c$ denote the level at which the unit size is $(c-1)\times (c-1)$ in our hierarchical grid scheme. Each core stores the variables for all Fourier modes $m$ at Level $l_c$ or below. This is not problematic, as the variable size remains small. The zero-boundary potential at Level 0 is computed using \cref{eq:matrix1}, and the D2D kernel operation in \cref{eq:d2d} can be performed in a bottom-up fashion, from Level $1$ to Level $l \leq l_c$, all within a single core, without communication between cores for all $m$. As mentioned in \autoref{sec:PSU}, the radial dependence of the kernels can be handled by the scale factor $\mathcal{R}_i$. Memory consumption is reduced by using shared memory across cores in the radial direction.

At Level $l>l_c$, the D2D kernel operation requires inter-core communication because $\rho_\text{mid}^l$ or $\sigma^l$ exceeds the grid range allocated to the core under the standard partitioning scheme. 
In this case, we partition the variables according to the mode number $m$: each core stores the variables for the full radial and polar ranges, but only for a specific range of $m$. This is achieved by using the \texttt{MPI\_scatter} function, which sends the boundary densities of the specified $m$-range at Level $l_c$ to the corresponding cores. Once this inter-core communication is complete, each core has the boundary densities of all units at Level $l_c$ across the entire $N_r \times N_\theta$ domain for the assigned $m$ range, which then serves as the input variable for the D2D kernel operation at Level $l_c+1$. This eliminates the need for further inter-core communication during the kernel operations at higher levels.

We apply the same partitioning scheme to the variables used in the D2P and P2P kernel operations, as well as to $\Phi_b^m$ required for the DGF. The P2P kernel operations follow a top-down approach, where the grid level for computation is progressively reduced. Once the level reaches $l=l_c$, the calculation can be performed within a single core again. The inter-core communication in this case occurs in the reverse direction compared to the previous bottom-up stage. The potential for a specific range of $m$, computed at the boundaries of the units at Level $l_c$ is transmitted back to the corresponding cores with the assigned radial and polar ranges. 

This method requires transmitting the density or potential at the boundaries of the grid assigned to a single core only twice. The corresponding communication size per core is $2\times 4cN_\phi = 8N_\phi(N_r N_\theta /N_\text{core})^{1/2}$, which is relatively small compared to the communication required for MHD variables.

\bibliography{Poisson}{}
\bibliographystyle{aasjournal}

\end{document}